\documentclass{iopart}
\usepackage{setstack,iopams}

\newtheorem{theorem}{Theorem}

\newtheorem{remark}[theorem]{Remark}

\begin{document}

\title{Towards a Theory of Chaos Explained as Travel on Riemann
Surfaces}

\author{F Calogero$^{1,2}$, D G\'omez-Ullate$^{3}$, P M Santini$^{1,2}$, and M Sommacal$^{4,5}$}
\address{$^1$\ Dipartimento di Fisica, Universit\`{a} di Roma ``La Sapienza'', Roma, Italy. }
\address{$^2$\ Istituto Nazionale di Fisica Nucleare, Sezione di Roma, Italy. }
\address{$^3$\ Departamento de F\'isica Te\'orica II, Universidad Complutense, Madrid, Spain.}
\address{$^4$\ Dipartimento di Matematica e Informatica, Universit\`{a} degli Studi di Perugia, Perugia, Italy. }
\address{$^5$\ Istituto Nazionale di Fisica Nucleare, Sezione di Perugia, Italy.  }
\eads{\mailto{francesco.calogero@roma1.infn.it},
  \mailto{david.gomez-ullate@fis.ucm.es}, \mailto{paolo.santini@roma1.infn.it}, \mailto{matteo.sommacal@pg.infn.it}}

\begin{abstract}
This paper presents a more complete version than hitherto published of our
explanation of a transition from \textit{regular} to \textit{irregular} motions and more
generally of the nature of a certain kind of \textit{deterministic chaos}.
To this end we introduced a simple model analogous to a
three-body problem in the plane, whose general solution is obtained via
\textit{quadratures} all performed in terms of \textit{elementary functions}.
For some values of the coupling constants the
system is \emph{isochronous} and explicit formulas for the period
of the solutions can be given. For other values, the motions are confined but feature
aperiodic (in some sense chaotic) motions.

This rich phenomenology can be understood in
remarkable, \textit{quantitative} detail in terms of travel on a
certain (circular) path on the Riemann surfaces defined by the
solutions of a related model considered as functions of a
\textit{complex} time. This model is meant to provide a
paradigmatic first step towards a somewhat novel understanding of a
certain kind of \textit{chaotic} phenomena.
\end{abstract}


 \maketitle

\section{Introduction}
The fact that the distinction among \textit{integrable} or \textit{%
nonintegrable} behaviors of a dynamical system is somehow connected with the
\textit{analytic structure} of the solutions of the model under
consideration as functions of the independent variable ``time'' (considered as a \emph{complex} variable) is by no means a
novel notion. It goes back to classical work by Carl Jacobi, Henri Poincar%
\'{e}, Sophia Kowalevskaya, Paul Painlev\'{e} and others. In recent times
some of us had the good fortune to hear in several occasions such ideas
clearly described by Martin Kruskal \cite{KC,kruskal}.
 A simple-minded rendition of his
teachings can be described as follows: for an evolution to be \textit{%
integrable}, it should be expressible, at least in principle, via formulas
that are \textit{not excessively multivalued} in terms of the dependent
variable, entailing that, to the extent this evolution is expressible by
\textit{analytic functions} of the dependent variable (considered as a
\textit{complex} variable), it might possess branch points, but it should
\textit{not} feature an \textit{infinity} of them that is \textit{dense} in
the \textit{complex} plane of the independent variable.

A number of techniques collectively known as Painlev\'e analysis have been
resurrected and further developed over the last few decades (for a review see, for
instance, \cite{RGB89}). In essence, they consider an \emph{ansatz} of the local
behaviour of a solution near a singularity in terms of a Laurent
series, introducing it in the equations and determining the
leading orders and resonances (terms in the expansion at which
arbitrary constants appear). Painlev\'e analysis has been extended
to test for the presence of algebraic branching (weak Painlev\'e
property \cite{RGB89}) by considering a Puiseaux series instead of
a Laurent series. These analytic techniques (which have been
algorithmized and are now available in computer packages)
constitute a useful tool in the investigation of integrability: in
many nonlinear systems where no solution in closed form is known,
Painlev\'e analysis provides information on the type of branching featured by
the general solution or by special classes of solutions. It has also
proved useful to identify special values of the parameters for
which generally \emph{chaotic} systems such as H\'enon-Heiles or Lorenz are
\emph{integrable} \cite{BSV,CTW82}.

On the opposite side of the spectrum lie chaotic dynamical systems
and it is natural to investigate the singularity structure of
their solutions. Tabor and his collaborators initiated this study
in the early eighties for the Lorenz system \cite{WT81} and the
Henon-Heiles Hamiltonian\cite{CTW82}. They realized that the
singularities of the solutions in complex-time are important for
the real-time evolution of the system. The complex time analytic
structure was studied by extensions of the Painlev\'e analysis
involving the introduction of logarithmic terms in the expansion
--- the so called $\Psi$-series --- which provides a local
representation of the solutions in the neighbourhood of a
singularity in the chaotic regime. Their local analytic approach
was complemented by numerical techniques developed for finding the
location of the singularities in complex time and determining the
order of  branching \cite{corliss}. In all the chaotic systems
under study, they observed numerically that the singularities in
complex time cluster on a natural boundary with self-similar
structure \cite{CGTW83}. An analytic argument to explain the
mechanism that leads to recursive singularity clustering was given
in \cite{LT88}. Similar studies relating singularity structure,
chaos and integrability have been performed by Bountis and his
collaborators. Going beyond the local techniques described above,
the emphasis is put on a global property of the solutions: whether
their Riemann surface has a finite or an infinite number of
sheets. Bountis proposes to use the term {\em integrable} for the
first case and {\em non-integrable} for the second \cite{BB,bount}.
Using mostly numerical evidence he conjectures that in the
non-integrable cases the Riemann surfaces are infinitely-sheeted
and the projection on the complex plane of the singularities is dense.
 Combining analytical and numerical results for a simple ODE, Bountis and
Fokas \cite{bountis_fokas} have identified chaotic systems with the
property that the singularities of their solutions are dense.

Painlev\'e analysis and its extensions are useful and widely
applicable. However, as local techniques, they  provide no
information on the {\em global} properties of the Riemann surfaces
of the solutions, such as: the number and location of the movable
branch points a solution has, and moreover how the different sheets of the
Riemann surface are connected together at those branch points.
Understanding these global properties is important for the
dynamics; a detailed analysis of the Riemann surface associated
to the solutions of a dynamical system, whenever it can be done,
provides a much deeper understanding than can be obtained by local
techniques alone.

This is precisely the motivation of the investigation reported herein: to introduce and study a model which is simple enough that a full description of its Riemann surface can be performed, yet complicated enough to feature a rich behaviour, possibly including irregular or chaotic characteristics.

Such a model was initially presented in \cite{CGSS2005} and in this paper we continue investigating its properties.
This line of research originates from a ``trick'' that is convenient to identify isochronous systems \cite{C1997,C2007b} -- a change of dependent and independent variables, with the new independent variable traveling on a path in the complex plane.  Later it was shown that many isochronous systems can
be written by a suitable modification of a large class of complex
ODEs \cite{CF02,C2007b}. Using local analysis and numerical integration
in two many-body systems in the plane \cite{CS2002,CFS2003}, it was
discovered that outside the isochrony region there exist periodic solutions with much higher periods as well as possibly aperiodic solutions, and the connection among this phenomenology and the analytic structure of the corresponding solutions as functions of complex time was illuminated.
However, those systems were too complicated for a complete description of the Riemann surface to be achieved.

Recent work along these lines includes problems whose solution is obtained by inversion of a hyperelliptic integral: the corresponding Riemann surfaces have been studied in \cite{FG2006,GS2006}, together with the implications on the dynamical properties of the models.

In the present paper we provide many details that were reported without proof in \cite{CGSS2005}, such as the description of the general solution by quadratures, and we also exhibit other properties of the model that were not present in \cite{CGSS2005}, such as similarity solutions, equilibrium configurations and small oscillations. Our investigation of this model will continue in a subsequent publication, \cite{CGSS2008}, where the full description of the geometrical properties of the Riemann surface will be given.

This paper is organized as follows: in  Section 2 we present our model, including in particular the relationship among its \emph{physical} version (independent variable: the \emph{real} time $t$) and its \emph{auxiliary} version (independent variable: a \emph{complex} variable $\tau$), and we outline the main findings reported in this paper. In Section \ref{Sec:equilibrium} we discuss the equilibrium configurations
of our \textit{physical} model, and the behavior of this system in the
neighborhood of these solutions, and we also obtain certain \textit{exact}
similarity solutions of our model and discuss their stability. In Section \ref{Sec:analytic} we discuss the analytic structure of the solutions of the \textit{auxiliary} model via local analyses \textit{\`{a} la Painlev\'{e}}, since the
analytic structure of these solutions plays a crucial role in determining
the time evolution of our \textit{physical} model.  In
Section \ref{Sec:General} we show how the \textit{general solution} of our model can be
achieved by \textit{quadratures} and in Section \ref{Sec:physical} we outline the behavior
of our model based on these results. Finally, in Section \ref{Sec:outlook} we summarize our results and comment on future developments.
 This paper also contains a few Appendices, where certain calculations are confined (to avoid interrupting
inconveniently the flow of the presentation) as well as certain additional
findings.

\section{Presentation of the model}

In this section we introduce the model treated in this paper, and we outline
our main findings that are then proven and further discussed in subsequent
sections.

\subsection{The auxiliary model}

The  \textit{auxiliary} model on which we focus in this paper is
characterized by the following system of \textit{three} coupled nonlinear
ODEs:%
\begin{equation}
\zeta _{n}^{\prime }=\frac{g_{n+2}}{\zeta _{n}-\zeta _{n+1}}+\frac{g_{n+1}}{%
\zeta _{n}-\zeta _{n+2}}~.  \label{1}
\end{equation}%
\textit{Notation}: here and hereafter indices such as $n,$ $m$ range from $1$
to $3$ and are defined $mod(3);$ $\tau $ is the (\textit{complex})
independent variable of this \textit{auxiliary} model; the \textit{three}
functions $\zeta _{n}\equiv \zeta _{n}(\tau )\ $are the dependent variables
of this \textit{auxiliary} model, and we assume them to be as well \textit{%
complex}; an appended prime always denotes differentiation with respect to
the argument of the functions it is appended to (here, of course, with
respect to the \textit{complex }variable $\tau )$; and the \textit{three}
quantities $g_{n}$ are arbitrary ``coupling constants'' (possibly also \textit{%
complex;} but in this paper we restrict consideration mainly to the case
with \textit{real} coupling constants; this is in particular hereafter
assumed in this section). In the following we will often focus on the
``semisymmetrical case'' characterized by the equality of \textit{two} of the
\textit{three }coupling constants, say%
\begin{equation}
g_{1}=g_{2}=g,~~~g_{3}=f, \label{Symm}
\end{equation}%
since in this case the treatment is simpler yet still adequate to exhibit
most aspects of the phenomenology we are interested in. More special cases
are the ``fully symmetrical'', or ``integrable'', one characterized by the
equality of \textit{all three} coupling constants,%
\begin{equation}
g=f,~~~g_{1}=g_{2}=g_{3}=g,  \label{Integr}
\end{equation}%
and the ``two-body'' one, with only one nonvanishing coupling constant, say
\numparts
\label{TwoBody}
\begin{equation}
g_{1}=g_{2}=g=0,~g_{3}=f\neq 0~.  \label{TwoBodya}
\end{equation}%
In this latter case clearly
\begin{equation}
\zeta _{3}^{\prime }=0,~\ ~\zeta _{3}(\tau )=\zeta _{3}(0)  \label{TwoBodyb}
\end{equation}%
(see (\ref{1})) and the remaining \textit{two-body} problem is trivially
solvable,%
\begin{equation}\fl\qquad
\zeta _{s}(\tau )=\frac{1}{2}\ \left[ \zeta _{1}(0)+\zeta _{2}(0)\right]
-(-)^{s}\,\left\{ \frac{1}{4}\ \left[ \zeta _{1}(0)-\zeta _{2}(0)\right]
^{2}+f\ \tau \right\} ^{\,1/2},~ s=1,2,  \label{TwoBodyc}
\end{equation}%
while the justification for labeling the fully symmetrical case (\ref{Integr}%
) as ``integrable'' will be clear from the following (or see Section 2.3.4.1\
of \cite{C2001}).
\endnumparts
Before introducing our \textit{physical} model, let us note that
the \textit{auxiliary }system (\ref{1}) is \textit{invariant} under
translations of both the independent variable $\tau $ (indeed, it is \textit{%
autonomous}) and the dependent variables $\zeta _{n}\left( \tau \right) $ ($%
\zeta _{n}\left( \tau \right) \Rightarrow \zeta _{n}\left( \tau \right)
+\zeta _{0},~\zeta _{0}^{\prime }=0$), and it is moreover \textit{invariant}
under an appropriate simultaneous rescaling of the independent and the
dependent variables.

\subsection{The trick and the physical model}\label{sec:trick}

The trick mentioned above, relating the \textit{auxiliary} model to the
\textit{physical} model, amounts in our present case to the introduction of
the (\textit{real}) independent variable $t$ (``physical time''), as well as
the \textit{three} (\textit{complex}) dependent variables $z_{n}\equiv
z_{n}(t)$, via the following positions:

\numparts\label{TRICK}
\begin{equation}
\tau =\frac{\exp (2\,i\,\omega \,t)-1}{2\,i\,\omega },  \label{TRICKb}
\end{equation}%
\begin{equation}
z_{n}(t)=\exp (-i\,\omega \,t)\,\zeta _{n}(\tau )~.  \label{TRICKa}
\end{equation}%
We hereafter assume the constant $\omega $ to be \textit{real }(for
definiteness, \textit{positive, }$\omega >0$; note that for $\omega =0$ the
change of variables disappears), and we associate to it the period
\begin{equation}\label{T}
T=\frac{\pi}{\omega}.
\end{equation}
Note that this change of variables entails that the initial values
$z_{n}(0)$ of the ``particle coordinates'' $z_{n}(t)$ coincide with the
initial values $\zeta _{n}(0)$ of the dependent variables of the \textit{%
auxiliary} model (\ref{1}):
\endnumparts
\begin{equation}
z_{n}(0)=\zeta _{n}(0)~.  \label{InDat}
\end{equation}

It is easily seen that, via this change of variables, (5), the
equations of motion (\ref{1}) satisfied by the quantities $\zeta _{n}(\tau )$
entail the following (\textit{autonomous}) equations of motion (in the
\textit{real} time $t$) for the particle coordinates $z_{n}(t)$:%
\begin{equation}
\dot{z}_{n}+i\,\omega \,z_{n}=\frac{g_{n+2}}{z_{n}-z_{n+1}}+\frac{g_{n+1}}{%
z_{n}-z_{n+2}}~.  \label{EqMot}
\end{equation}%
Here and hereafter superimposed dots indicate differentiations with respect
to the time $t.$

So, this model (\ref{EqMot}) describes the ``physical evolution'' which we
study. Note that its equations of motion, (\ref{EqMot}), are of \textit{%
Aristotelian}, rather than \textit{Newtonian}, type: the ``velocities'' $\dot{z%
}_{n}$, rather than the ``accelerations'' $\ddot{z}_{n}$, of the
moving particles are determined by the ``forces''. In Appendix D we
discuss the connection of this model with more classical many-body
problems, characterized by \textit{Newtonian} equations of motion.

Let us immediately emphasize two important \textit{qualitative} aspects of
the dynamics of our \textit{physical} model (\ref{EqMot}). The ``one-body
force'' represented by the second term, $i\,\omega \,z_{n}$, in the left-hand
side of the equations of motion (\ref{EqMot}) becomes dominant with respect
to the ``two-body forces'' appearing in the right-hand side in determining the
dynamics whenever the (\textit{complex}) coordinate $z_{n}$ of the $n$-th
particle becomes large (in modulus). Hence when $\left\vert
z_{n}(t)\right\vert $ is very large, the solution $z_{n}(t)$ of (\ref
{EqMot}) is approximated by the solution of $\dot{z}_{n}+i\,\omega
\,z_{n}\approx 0$ implying that $z_{n}\left( t\right) $ is characterized by
the behavior $z_{n}(t)\approx c\,\exp \left( -i\,\omega \,t\right) $,
therefore the trajectory of the $n$-th particle tends to rotate (clockwise,
with period $2\,T$) on a (large) circle. This effect causes \textit{all }%
motions of our \textit{physical} model, (\ref{EqMot}), to be \textit{confined%
}. Secondly, it is clear that the \textit{two-body} forces (see the
right-hand side of (\ref{EqMot})) cause a \textit{singularity} whenever
there is a \textit{collision} of \textit{two} (or all \textit{three}) of the
particles as they move in the \textit{complex} $z$-plane, and become
dominant whenever \textit{two} or \textit{three} particles get very close to
each other, namely in the case of \textit{near collisions}. But if the \textit{%
three} particles move \textit{aperiodically} in a \textit{confined} region
(near the origin) of the complex $z$-plane, a lot of \textit{near collisions}
shall indeed occur. And since the outcome of a \textit{near collision} is likely
to be quite different depending on which side two particles scatter past
each other -- and this, especially in the case of very close \textit{near collisions}, depends \textit{sensitively} on the initial data of the trajectory
under consideration -- we see here a mechanism complicating the motion,
indeed causing some kind of \textit{chaos} associated with a \textit{%
sensitive dependence} of the motion on its initial data. This suggests that
our model (\ref{EqMot}), in spite of its simplicity, is likely to be rich
enough to cause an interesting dynamical evolution. We will see that this is
indeed the case. But before proceeding with this investigation let us
interject two remarks (somewhat related to each other).

\begin{remark} This system (\ref{EqMot}) is still invariant under
translations of the independent variable $t$ (indeed, it is again \emph{%
autonomous}) but, in contrast to (\ref{1}), it is no longer invariant under
translations of the dependent variables $z_{n}(t)$ nor under a simple
rescaling of the independent variable $t$ and of the dependent variables $%
z_{n}(t)$.
\end{remark}

\begin{remark} The \textit{general solution} of the equations of motion (%
\ref{EqMot}) has the structure
\numparts
\label{CMStru}
\begin{equation}
z_{n}(t)=z_{CM}(t)+\check{z}_{n}(t),  \label{CMStrua}
\end{equation}%
where the \textit{three} functions $\check{z}_{n}(t)$ satisfy themselves the
same equations of motion (\ref{EqMot}) as well as the additional restriction%
\begin{equation}
\check{z}_{1}(t)+\check{z}_{2}(t)+\check{z}_{3}(t)=0~  \label{CMStrub}
\end{equation}%
which is clearly compatible with these equations of motion, and
correspondingly $z_{CM}(t)$ is the center of mass of the system (\ref{EqMot}%
),
\endnumparts
\numparts
\label{CM}
\begin{equation}
z_{CM}(t)=\frac{z_{1}(t)+z_{2}(t)+z_{3}(t)}{3},  \label{CMa}
\end{equation}%
and it evolves according to the simple formula
\begin{equation}
z_{CM}(t)=z_{CM}(0)\,\exp \left( -i\,\omega \,t\right) =Z\,\exp \left(
-i\,\omega \,t\right) ~. \label{CMb}
\end{equation}\endnumparts
\end{remark}

In Section \ref{Sec:equilibrium} (and Appendix A) we determine the equilibrium configurations of
our \textit{physical} model, namely the values $z^{\mbox{\tiny(eq)}}_{n}$ of the \textit{%
three} particle coordinates $z_{n}$ such that
\begin{equation}
z_{n}=z^{\mbox{\tiny(eq)}}_{n},~~~\dot{z}_{n}=0  \label{EqConfig}
\end{equation}%
satisfy the equations of motion (\ref{EqMot}), and we ascertain the behavior
of our system in the neighborhood of these configurations. In the second part of
Section \ref{Sec:equilibrium}, and then almost always in the rest of this paper (and throughout
the rest of this section) we restrict for simplicity consideration to the
semisymmetrical case, see (\ref{Symm}). A main finding in Section \ref{Sec:equilibrium} (and
Appendix A) is that in the semisymmetrical case our model (\ref{EqMot})
possesses generally \textit{two} equilibrium configurations $z^{\mbox{\tiny(eq)}}_{n}$.
We moreover determine the \textit{three} exponents $\gamma ^{(m)}$
characterizing the small oscillations of our system in the neighborhood of
each of these \textit{two }configurations, defined according to the standard
formulas (see Section \ref{Sec:equilibrium})
\numparts
\begin{equation}
z_{n}(t)=z^{\mbox{\tiny(eq)}}_{n}+\varepsilon \,w_{n}(t),  \label{NearEqa}
\end{equation}%
\begin{equation}
w_{n}^{(m)}(t)=\exp (-i\,\gamma ^{(m)}\,\omega \,t)\,v_{n}^{(m)},
\label{NearEqb}
\end{equation}%
where of course $\varepsilon $ is an \emph{infinitesimally small} parameter and the
quantities $v_{n}^{(m)}$ are time-independent. We find that the first
\textit{two} of these\textit{\ three} exponents take in \textit{both} cases
the simple values
\endnumparts
\begin{equation}
\gamma ^{(1)}=1,~~\ \gamma ^{(2)}=2~;
\end{equation}%
the first of these corresponds of course to the center-of-mass motion, see (%
\ref{CMb}). As for the \textit{third} exponent $\gamma ^{(3)},$ we find for
one equilibrium configuration
\numparts
\label{GAMMA3}
\begin{equation}
\gamma ^{(3)}=\frac{f+8\ g}{f+2\!g}=\frac{1}{%
\mu },  \label{GAMMA3a}
\end{equation}%
and for the other%
\begin{equation}
\gamma ^{(3)}=\frac{f+8\,g}{3\,g}=\frac{2}{1-\mu }~. \label{GAMMA3b}
\end{equation}%
Here we have introduced the
constant $\mu ,$%
\endnumparts
\begin{equation}
\mu =\frac{f+2\,g}{f+8\,g} \label{mu}
\end{equation}%
whose value, as we shall see, plays an important role in determining the
dynamical evolution of our model: in particular, this evolution does largely
depend on whether or not $\mu $ is a \textit{real rational} number, and if
it is \textit{rational},%
\begin{equation}
\mu =\frac{p}{q}  \label{mupq}
\end{equation}%
with $p$ and $q$ \textit{coprime integers} (and $q$ \textit{positive}, $q>0$%
), on whether the two natural numbers $\left\vert p\right\vert $ and $q$ are
large or small. A hint of this is already apparent from the results we just
reported: while the solutions $w_{n}^{(1)}(t)$ and $w_{n}^{(2)}(t),$ see (%
11), are both \textit{periodic} with period $2\,T$ \thinspace (see
(\ref{T}); in fact $w_{n}^{(2)}(t)$ is periodic with period $T$), the
solution $w_{n}^{(3)}(t),$ see (11), is periodic with the period $%
\tilde{T},$%
\begin{equation}
\tilde{T}=\frac{2\,T}{\gamma ^{(3)}},  \label{Ttilde}
\end{equation}%
which is clearly \textit{congruent} to $T$\textit{only }if $\mu $ is \textit{rational}, see (\ref
{mupq}) and (13) -- implying then that the small
oscillations around the equilibrium configurations are always \textit{%
completely periodic} with a period which is a \textit{finite
integer} multiple of $T$.

In Section \ref{Sec:equilibrium} we also introduce the special class of (\textit{exact} and
\textit{completely explicit}) ``similarity'' solutions of our equations of
motion, (\ref{EqMot}), and analyze their \textit{stability}, namely the
solutions of our system in the immediate neighborhood of these similarity
solutions.

\subsection{Conserved quantities}

It is important to note at this point that the auxiliary model (\ref{1}) possesses conserved quantities, which will be used in Section \ref{Sec:General} to obtain its general solution by quadratures.
Firstly, due to the translational invariance it is obvious that the quantity
\begin{equation}
Z=\frac{1}{3}\sum\limits_{n=1}^{3}\zeta _{n},
\end{equation}%
does not depend on $\tau$.
In addition, the analysis of Section \ref{Sec:General} shows that in the semisymmetrical case there exists an extra conserved quantity given by
\begin{equation}\fl\qquad
\tilde{K}=\left( 2\,\zeta _{3}-\zeta _{1}-\zeta _{2}\right) ^{-2}\,\left[ 1-%
\frac{\left( \zeta _{1}-\zeta _{2}\right) ^{2}+\left( \zeta _{2}-\zeta
_{3}\right) ^{2}+\left( \zeta _{3}-\zeta _{1}\right) ^{2}}{2\,\mu \,\left(
2\,\zeta _{3}-\zeta _{1}-\zeta _{2}\right) ^{\,2}}\right] ^{\,\mu -1}~.
\label{Ktilde}
\end{equation}%
Here the constant $\mu $ is defined in terms of the
coupling constants $g$ and $f,$ see (\ref{Symm}), by (\ref{mu}). We already
mentioned that the value of this parameter (in particular, whether or not $%
\mu $ is a \textit{rational} number) plays an important role in determining
the dynamical evolution of our model. A hint of this is now provided
by the appearance of this number $\mu $ as an exponent in the right-hand
side of (\ref{Ktilde}), since this exponent characterizes the
multivaluedness of the dependence of the constant $\tilde{K}$ on the
 coordinates $\zeta _{n}$.

\section{Equilibrium configurations, small oscillations and similarity solutions of the \textit{physical} model}\label{Sec:equilibrium}

In this section we discuss, firstly, the equilibrium configurations of our
\textit{physical }model, (\ref{EqMot}), and its behavior near equilibrium,
and secondly, a special, explicit ``similarity'' solution of our model and its
stability.

The \textit{equilibrium configurations} of our \textit{physical }model (\ref
{EqMot}),
\begin{equation}
z_{n}(t)=z^{\mbox{\tiny(eq)}}_{n},~\ \ \dot{z}_{n}(t)=0,  \label{EqConf}
\end{equation}%
(see (\ref{EqConfig})) are clearly characterized by the algebraic equations%
\begin{equation}
i\,\omega \,z^{\mbox{\tiny(eq)}}_{n}=\frac{g_{n+1}}{z^{\mbox{\tiny(eq)}}_{n}-z^{\mbox{\tiny(eq)}}_{n+2}}+\frac{%
g_{n+2}}{z^{\mbox{\tiny(eq)}}_{n}-z^{\mbox{\tiny(eq)}}_{n+1}}~.  \label{EqEq}
\end{equation}
These algebraic equations entail%
\begin{equation}
z^{\mbox{\tiny(eq)}}_{1}+z^{\mbox{\tiny(eq)}}_{2}+z^{\mbox{\tiny(eq)}}_{3}=0~.  \label{SumZero}
\end{equation}
It is now convenient  to set%
\begin{equation}
z^{\mbox{\tiny(eq)}}_{n}=\left( 2\ i\ \omega \right) ^{-1/2}\ \alpha _{n},  \label{alpha}
\end{equation}%
so that the equilibrium equations (\ref{EqEq}) read as follows:%
\begin{equation}
\frac{\alpha _{n}}{2}=\frac{g_{n+1}}{\alpha _{n}-\alpha _{n+2}}+\frac{g_{n+2}%
}{\alpha _{n}-\alpha _{n+1}}~.  \label{Ansbb}
\end{equation}%
These algebraic equations can be conveniently (see below) rewritten as
follows:
\numparts
\begin{equation}
\alpha _{n}=\beta _{n+1}\,\left( \alpha _{n}-\alpha _{n+2}\right) +\beta
_{n+2}\,\left( \alpha _{n}-\alpha _{n+1}\right) ,  \label{600a}
\end{equation}%
via the position%
\begin{equation}
\beta _{n}=\frac{2\,g_{n}}{\left( \alpha _{n-1}-\alpha _{n+1}\right) ^{2}}~.
\label{600b}
\end{equation}
We now note that, in order that the three equations (\ref{600a}) (which are
linear in the three unknowns $\alpha _{n}$, although only apparently so, see
(\ref{600b})) have a nonvanishing solution, the quantities $\beta _{n}$ must
cause the following determinant to vanish:
\endnumparts
\begin{equation}
\left\vert
\begin{array}{ccc}
\beta _{2}+\beta _{3}-1 & -\beta _{3} & -\beta _{2} \\
-\beta _{3} & \beta _{3}+\beta _{1}-1 & -\beta _{1} \\
-\beta _{2} & -\beta _{1} & \beta _{1}+\beta _{2}-1%
\end{array}%
\right\vert =0~.  \label{601}
\end{equation}

To analyze the \textit{small oscillations} of our system (\ref{EqMot})
around its equilibrium configurations we now set
\numparts
\label{w}
\begin{equation}
z_{n}(t)=z^{\mbox{\tiny(eq)}}_{n}+\varepsilon \,w_{n}(t),  \label{wa}
\end{equation}%
(see (\ref{NearEqa})) and we then get (linearizing by treating $\varepsilon $
as an infinitesimally  small parameter)%
\begin{equation}\fl\qquad\qquad
\dot{w}_{n}+i\,\omega \,w_{n}+i\,\omega \,\beta _{n+1}\,\left(
w_{n}-w_{n+2}\right) +\beta _{n+2}\,\left( w_{n}-w_{n+1}\right) =0~.
\label{wb}
\end{equation}%
Therefore the three exponents $\gamma ^{(m)}$ characterizing the small
oscillations around equilibrium via the formula
\endnumparts
\begin{equation}
w_{n}^{(m)}(t)=\exp (-i\,\gamma ^{(m)}\,\omega \,t)\,v_{n}^{(m)},
\label{wc}
\end{equation}%
providing \textit{three} independent solutions of the system of \textit{%
linear} ODEs (\ref{wb}), are the \textit{three} eigenvalues of the
symmetrical matrix%
\begin{equation}
\mathbf{B}=\left(
\begin{array}{ccc}
\beta _{2}+\beta _{3}+1 & -\beta _{3} & -\beta _{2} \\
-\beta _{3} & \beta _{3}+\beta _{1}+1 & -\beta _{1} \\
-\beta _{2} & -\beta _{1} & \beta _{1}+\beta _{2}+1%
\end{array}%
\right) ,  \label{MatrixB}
\end{equation}%
and the \textit{three} $3$-vectors $\vec{v}^{(m)}\equiv \left(
v_{1}^{(m)},v_{2}^{(m)},v_{3}^{(m)}\right) $ are the corresponding
eigenvectors,%
\begin{equation}
\sum_{\ell =1}^{3}B_{n\ell }\,v_{\ell }^{(m)}=\gamma ^{(m)}\,v_{n}^{(m)}~.
\end{equation}%
Hence the \textit{three} exponents $\gamma ^{(m)}$ are the \textit{three}
roots of the ``secular equation''
\begin{equation}
\left\vert
\begin{array}{ccc}
\beta _{2}+\beta _{3}+1-\gamma & -\beta _{3} & -\beta _{2} \\
-\beta _{3} & \beta _{3}+\beta _{1}+1-\gamma & -\beta _{1} \\
-\beta _{2} & -\beta _{1} & \beta _{1}+\beta _{2}+1-\gamma%
\end{array}%
\right\vert =0~.  \label{Detbeta}
\end{equation}%
Clearly these \textit{three} roots are given by the following formulas:%
\begin{equation}
\gamma ^{(1)}=1,~~~\gamma ^{(2)}=2,~~~\gamma ^{(3)}=2\!\left( \beta
_{1}+\beta _{2}+\beta _{3}\right) ~.  \label{3gammas}
\end{equation}%
Indeed the determinant (\ref{Detbeta}) vanishes for $\gamma =\gamma ^{(1)}=1$
(when each line sums to zero) and for $\gamma =\gamma ^{(2)}=2$ (see (\ref
{601})), and the third solution,%
\begin{equation}
\gamma ^{(3)}=2\!\left( \beta _{1}+\beta _{2}+\beta _{3}\right) ,
\label{gamma3}
\end{equation}%
is then implied by the trace condition%
\begin{equation}
\mbox{trace}\left[ \mathbf{B}\right] =3+2\!\left( \beta _{1}+\beta
_{2}+\beta _{3}\right) =\gamma ^{(1)}+\gamma ^{(2)}+\gamma ^{(3)}~.
\end{equation}%
The first of these $3$ solutions, $\gamma ^{(1)}=1,$ corresponds to the
center of mass motion (it clearly entails $v_{n}^{(1)}=v^{(1)},$ see (\ref
{wc}) and (\ref{MatrixB})).

In the semisymmetrical case (\ref{Symm}) the equations (\ref{Ansbb}) (or
equivalently (24)) characterizing, via (\ref{alpha}), the equilibrium
configurations can be solved explicitly (see Appendix A). One finds that
there are \textit{two} distinct equilibrium configurations (in fact four, if
one takes account of the trivial possibility to exchange the roles of the
two ``equal'' particles with labels $1$ and $2),$ the first of which reads
simply%
\begin{equation}\fl\qquad\qquad
z^{\mbox{\tiny(eq)}}_{3}=0,\qquad z^{\mbox{\tiny(eq)}}_{1}=-z^{\mbox{\tiny(eq)}}_{2}=z^{\mbox{\tiny(eq)}},\qquad\left( z^{\mbox{\tiny(eq)}}\right)
^{\,2}=\frac{f+2\,g}{2\,i\,\omega },  \label{FirstEqConf}
\end{equation}%
while the second has a slightly more complicated expression (see Appendix
A). Note however that, \textit{in both cases}, there holds the relation%
\begin{equation}\fl\qquad\qquad
(z^{\mbox{\tiny(eq)}}_{1}-z^{\mbox{\tiny(eq)}}_{2})^{2}+(z^{\mbox{\tiny(eq)}}_{2}-z^{\mbox{\tiny(eq)}}_{3})^{2}+(z^{\mbox{\tiny(eq)}}_{3}-%
z^{\mbox{\tiny(eq)}}_{1})^{2}=\frac{3\,\left( f+2\,g\right) }{i\,\omega }~.  \label{EQEQ}
\end{equation}%
Moreover, in both cases the corresponding values for the eigenvalue $\gamma
^{(3)},$ see (\ref{gamma3}), are easily evaluated. The first solution yields
(see (\ref{GAMMA3a}))
\numparts
\label{Gamma3}
\begin{equation}
\gamma ^{(3)}=\frac{f+8\ g}{f+2\!\,g}=\frac{1}{%
\mu }=\frac{q}{p},  \label{Gamma3a}
\end{equation}%
where, for future reference, we expressed $\gamma ^{(3)}$ not only in terms
of the parameter $\mu $, see (\ref{mu}), but as well in terms of its
rational expression (\ref{mupq}) (whenever applicable), while the second
solution likewise yields%
\begin{equation}
\gamma ^{(3)}=\frac{f+8\!g}{3\!\,g}=\frac{2}{1-\mu }=
\frac{2\,q}{q-p}~.  \label{Gamma3b}
\end{equation}%
Note that this implies that in the ``integrable'' case (\ref{Integr}) both these formulas, (\ref{Gamma3a}) and (\ref{Gamma3b}),
yield $\gamma ^{(3)}=3$; but it is easily seen that in this case only the
first equilibrium configuration (\ref{FirstEqConf}) actually exists. So in
the ``integrable'' case the oscillations around the (only) equilibrium
configuration (\ref{FirstEqConf}) are the linear superposition of three
periodic motions (see (\ref{wc})) with respective periods $2\,T,$ $T$ and $%
\frac{2\,T}{3}$ (see (\ref{T})).\ Also in the ``two-body'' case (4) the second equilibrium configuration does not exist, while the first
formula, (\ref{Gamma3a}), yields $\gamma ^{(3)}=1,$ so in this case the
small oscillations around the (only) equilibrium configuration (\ref
{FirstEqConf}) are the linear superposition of two periodic motions, with
periods $2\,T$ and $T$ (see (\ref{T}); consistently with the explicit
solution, easily obtainable from (\ref{TwoBodyc}) via (5)).

As can be easily verified, the equilibrium configurations (\ref{EqConf})
with (\ref{EqEq}) are merely the special case corresponding to $%
z_{CM}(0)=0,~c=0$ of the following two-parameter family of (\textit{exact})
``similarity'' solutions of our equations of motion (\ref{EqMot}):
\endnumparts
\numparts
\begin{equation}
z_{n}(t)=z_{CM}\,(t)+\tilde{z}_{n}(t;c),  \label{SimSolZa}
\end{equation}%
\begin{equation}
\tilde{z}_{n}(t;c)\equiv z^{\mbox{\tiny(eq)}}_{n}\,\left[ 1+c\,\exp \left( -2\,i\,\omega
\,t\right) \right] ^{1/2},  \label{SimSolZb}
\end{equation}%
with the center of mass coordinate $z_{CM}(t)$ evolving according to (\ref
{CMb}). The two \textit{arbitrary }(\textit{complex) }constants featured by
this solution are of course $z_{CM}(0)=Z$ (see (\ref{CMb})) and $c$, while
the constants $z^{\mbox{\tiny(eq)}}_{n}$'s are defined as in the preceding section, see (%
\ref{EqEq}).

Clearly these (\textit{exact}) solutions correspond, via the trick (5), the relation (\ref{alpha}) (which is clearly consistent with (\ref{Ansbb}) and (\ref{EqEq})) and the simple relation
\endnumparts
\begin{equation}
\tau _{b}=\frac{c-1}{2\ i\ \omega },
\end{equation}%
to the two-parameter family
\begin{equation}
\zeta _{n}(\tau )=Z+\alpha _{n}\,\left( \tau -\tau _{b}\right) ^{1/2},
\label{exact}
\end{equation}%
of (\textit{exact}) solutions of (\ref{1}).


Let us now discuss the stability of this solution, (\ref{SimSolZb}). To this end we set
\numparts
\label{wtil}
\begin{equation}
z_{n}(t)=\tilde{z}_{n}(t;c)+\varepsilon \,\tilde{w}_{n}(t),  \label{wtila}
\end{equation}%
and we insert this \textit{ansatz} in our equations of motion (\ref{EqMot}),
linearizing them by treating $\varepsilon $ as an infinitesimally small parameter. We thus get%
\begin{equation}\fl\qquad\qquad
\overset{\cdot }{\tilde{w}}_{n}+i\,\omega \,\tilde{w}_{n}+\frac{i\,\omega \,%
\left[ \beta _{n+1}\,\left( \tilde{w}_{n}-\tilde{w}_{n+2}\right) +\beta
_{n+2}\,\left( \tilde{w}_{n}-\tilde{w}_{n+1}\right) \right] }{1+c\,\exp
\left( -2\,i\,\omega \,t\right) }=0,  \label{wtilb}
\end{equation}%
having used the definition (\ref{600b}).
Clearly the solution of this system of ODEs reads
\endnumparts
\numparts
\label{xhi}
\begin{equation}
\tilde{w}_{n}(t)=\exp \left( -i\,\omega \,t\right) \mathit{\,}\chi
_{n}\left( \vartheta \right) ,  \label{chia}
\end{equation}%
with%
\begin{equation}
\vartheta \equiv \vartheta (t)=t-\left( 2\,i\,\omega \right) ^{-1}\,\log
\left[ \frac{1+c\,\exp \left( -2\,i\,\omega \,t\right) }{1+c}\right]
\label{chib}
\end{equation}%
and the functions $\chi _{n}\left( \vartheta \right) $ solutions of the autonomous linear system of \textit{first-order} ODEs%
\begin{equation}
\chi _{n}^{\prime }+i\,\omega \,\left[ \beta _{n+1}\,\left( \chi_{n}-%
\chi_{n+2}\right) +\beta _{n+2}\,\left( \chi_{n}-\chi
_{n+1}\right) \right] =0,  \label{chic}
\end{equation}%
where the primes denote of course differentiation with respect to $\vartheta
.$ Hence (see (\ref{wb})) the \textit{three} independent solutions of this
linear system are%
\begin{equation}
\chi _{n}^{(m)}\left( \vartheta \right) =\exp \left( i\,\omega \,\vartheta
\right) \,w_{n}^{(m)}\left( \vartheta \right) ,  \label{chid}
\end{equation}%
with the functions $w_{n}^{(m)}$ defined by (\ref{wc}) (of course with $t$
replaced by $\vartheta $), yielding via (\ref{wc}) and (\ref{chia}) with (%
\ref{chib}) the following two equivalent expressions for the \textit{three}
independent solutions of the linear system (\ref{wtilb}):
\endnumparts
\numparts
\label{wti}
\begin{equation}\fl\qquad\qquad
\tilde{w}_{n}^{(m)}(t)=\left[ \frac{1+c\,\exp \left( -2\,i\,\omega
\,t\right) }{1+c}\right] ^{(\gamma ^{(m)}-1)/2}\,\exp (i\,\gamma
^{(m)}\,\omega \,t)\,\tilde{v}_{n}^{(m)},  \label{wtia}
\end{equation}%
\begin{equation}\fl\qquad\qquad
\tilde{w}_{n}^{(m)}(t)=\left[ \frac{\exp \left( 2\,i\,\omega \,t\right) +c\,%
}{1+c}\right] ^{(\gamma ^{(m)}-1)/2}\,\exp (i\,\omega \,t)\,\tilde{v}%
_{n}^{(m)}~.  \label{wtic}
\end{equation}%
Here the \textit{three} exponents $\gamma ^{(m)}$ are defined as above, see (%
\ref{3gammas}), and likewise the ``eigenvectors'' $\tilde{v}_{n}^{(m)}$
coincide with those defined above up to (\textit{arbitrary}) normalization
constants $c^{(m)}$,%
\begin{equation}
\tilde{v}_{n}^{(m)}=c^{(m)}\,v_{n}^{(m)}~.  \label{wtid}
\end{equation}%
Note the equivalence of the two expressions (\ref{wtia}) and (\ref{wtic})
(the motivation for writing these two versions of the same formula will be
immediately clear).

For $m=1,2,3$ these solutions, see (\ref{wtic}), are periodic functions of
the (\textit{real}) time $t$ with period $2\,T$ if $\left\vert c\right\vert
>1.$ If instead $\left\vert c\right\vert <1$, the solutions (see (\ref{wtia}%
)) with $m=1$ respectively $m=2$ are periodic with periods $2\,T$
respectively $T$ (see (\ref{3gammas})). The solution with $m=3$ is \textit{periodic} if $\gamma
^{(3)}$ is \textit{real, }but with the period $\frac{2\,T}{\left\vert \gamma
^{(3)}\right\vert }$ which is not congruent to $T$ if $\gamma ^{(3)}$ is
\textit{irrational}; it grows exponentially with increasing time if $\mbox{Im%
}\left[ \gamma ^{(3)}\right] <0$, implying instability of the solution (37) in this case, and it instead decays exponentially if $\mbox{Im}%
\left[ \gamma ^{(3)}\right] >0$, implying a limit cycle behavior in
configuration space, namely asymptotic approach to a solution \textit{%
completely periodic} with period $T$ or $2\,T$ depending whether the center
of mass of the system is fixed at the origin or itself moving with period $%
2\,T$; but note that in this paper we restrict our attention to the case with \textit{real}
coupling constants.


\bigskip

\section{Analytic structure of the solutions of the \textit{auxiliary} model}\label{Sec:analytic}

In this section we discuss the
properties of analyticity as functions of the complex variable $\tau $ of
the solutions $\zeta _{n}(\tau )$ of the \textit{auxiliary} model (\ref{1})
(with arbitrary values of the $3$ coupling constants $g_{n}$, i. e. \textit{%
not} restricted by the semisymmetrical condition (\ref{Symm}): except when
this is explicitly specified, see below). In particular we show first of all
that, for appropriate initial data characterized by \textit{sufficiently
large} values of the moduli of \textit{all three} interparticle distances,
namely by the condition (see (\ref{InDat})) that the quantity
\endnumparts
\begin{equation}
\zeta _{\min }=\underset{n,m=1,2,3;~n\neq m}{\min }\left\vert \zeta
_{n}(0)-\zeta _{m}(0)\right\vert  \label{zimin}
\end{equation}%
be \textit{adequately large}, the solutions $\zeta _{n}(\tau )$ are \textit{%
holomorphic} in a disk $D_{0}$ of (\textit{arbitrarily large}) radius $d_{0}$
centered at the origin, $\tau =0,$ of the complex $\tau $-plane (of course
the ``adequately large'' value of the quantity $\zeta _{\min }$ depends on $%
d_{0},$ and on the magnitude of the \textit{three }coupling constants $g_{n}$%
; see (\ref{dzeroupper}) below). We moreover discuss via a local analysis
\textit{a la Painlev\'{e} }the nature of the singularities of the solutions $%
\zeta _{n}(\tau )$ of the \textit{auxiliary} model (\ref{1}) as functions of
the \textit{complex} variable $\tau $ and we thereby justify the assertions
made in this respect in Section 2.

To prove the first point, set
\numparts
\label{Chi}
\begin{equation}
\sigma _{n}(\tau )=\zeta _{n}(\tau )-\zeta _{n}(0),  \label{Chia}
\end{equation}%
so that these quantities $\sigma _{n}(\tau )$ vanish initially,%
\begin{equation}
\sigma _{n}(0)=0,  \label{Chib}
\end{equation}%
and, as a consequence of (\ref{1}), satisfy the equations of motion

\begin{eqnarray}
\sigma _{n}^{\prime }(\tau ) &=&\frac{g_{n+1}}{\zeta _{n}(0)-\zeta
_{n+2}(0)+\sigma _{n}(\tau )-\sigma _{n+2}(\tau )}  \nonumber \\
&&+\frac{g_{n+2}}{\zeta _{n}(0)-\zeta _{n+1}(0)+\sigma _{n}(\tau
)-\sigma _{n+1}(\tau )}~.  \label{Chic}
\end{eqnarray}%
A standard theorem (see, for instance, \cite{Ince}) guarantees then that
these quantities $\sigma _{n}(\tau )$ -- hence as well the functions $%
\zeta _{n}(\tau ),$ see (\ref{Chia}) -- are \textit{holomorphic} in $\tau $
(at least) in a disk $D_{0}$ centered at the origin $\tau =0$ in the \textit{%
complex }$\tau $-plane, the radius $d_{0}$ of which is bounded \textit{below}
by the inequality
\endnumparts
\begin{equation}
d_{0}>\frac{b}{4\,M}  \label{dzero}
\end{equation}%
(this formula coincides with the last equation of Section 13.21 of \cite%
{Ince}, with the assignments $m=3$ and $a=\infty ,$ the first of which is
justified by the fact that the system (\ref{Chic}) features $3$ coupled
equations, the second of which is justified by the \textit{autonomous}
character of the equations of motion (\ref{Chic})). The two \textit{positive}
quantities $b$ and $M$ in the right-hand side of this inequality are defined
as follows. The quantity $b$ is required to guarantee that the right-hand
sides of the equations of motion (\ref{Chic}) be \textit{holomorphic} (as
functions of the dependent variables $\sigma _{n})$ provided these
quantities satisfy the three inequalities%
\begin{equation}
\left\vert \sigma _{n}\right\vert \leq b~;  \label{b}
\end{equation}%
clearly in our case a sufficient condition to guarantee this is provided by
the single restriction
\begin{equation}
b<\frac{\zeta _{\min }}{2},  \label{B}
\end{equation}%
with $\zeta _{\min }$ defined by (\ref{zimin}). The second quantity in the
right-hand side of (\ref{dzero}), $M\equiv M(b),$ is the \textit{upper}
bound of the right-hand sides of (\ref{Chic}) when the quantities $\sigma
_{n}$ satisfy the inequality (\ref{b}); but of course the inequality (\ref{dzero}) holds \textit{a fortiori }if we overestimate $M,$ as we shall
presently do. Indeed clearly the equations of motion (\ref{Chic}) with (\ref{b}) and (\ref{B}) entail%
\begin{equation}
M<\frac{4\,G}{\zeta _{\min }-2\,b},  \label{MaxM}
\end{equation}%
with%
\begin{equation}
G=\underset{n=1,2,3}{\max }\left\vert g_{n}\right\vert ~.
\end{equation}%
Insertion of (\ref{MaxM}) in (\ref{dzero}) yields%
\begin{equation}
d_{0}>\frac{b\,\left( \zeta _{\min }-2\,b\right) }{16\,G},
\end{equation}%
hence, setting $b=\frac{\zeta _{\min }}{4}$ (to maximize the right-hand
side; note the consistency of this assignment with (\ref{B}))$,$%
\begin{equation}
d_{0}>\frac{\zeta _{\min }^{2}}{128\,G},  \label{dzeroupper}
\end{equation}%
confirming the assertion made above (that $d_{0}$ can be made \textit{%
arbitrarily large} by choosing $\zeta _{\min }$ \textit{adequately large}).

Next, let us show, via a local analysis \textit{\`{a} la Painlev\'{e}}, that
the singularities as functions of the complex variable $\tau $ of the
\textit{general }solutions $\zeta _{n}(\tau )$ of our \textit{auxiliary}
model (\ref{1}) associated with a coincidence of \textit{two} of the \textit{%
three} components $\zeta _{n}$ are \textit{square-root} \textit{branch points%
} (recall that a singularity at finite $\tau $ of a solution $\zeta
_{n}(\tau )$ of the evolution equations (\ref{1}) may only occur when the
right-hand side of these equations diverges).\textit{\ }Such a singularity
occurs for those values $\tau _{b}$ of the independent variable $\tau $ such
that \textit{two} of the \textit{three} functions $\zeta _{n}$ coincide, for
instance%
\begin{equation}
\zeta _{1}(\tau _{b})=\zeta _{2}(\tau _{b})\neq \zeta _{3}(\tau _{b})~.
\label{TwoBodyColl}
\end{equation}

The \textit{square-root} character of these branch points is evident from
the following \textit{ansatz} characterizing the behavior of the solutions
of (\ref{1}) in the neighborhood of these singularities:
\numparts
\begin{equation*}\fl\qquad
\zeta _{s}(\tau )=\zeta _{b}-(-1)^{s}\ \alpha \,\left( \tau -\tau
_{b}\right) ^{1/2}+v_{s}\,\left( \tau -\tau _{b}\right)
+\sum\limits_{k=3}^{\infty }\alpha _{s}^{(k)}\,\left( \tau -\tau _{b}\right)
^{k/2},\quad s=1,2 
\end{equation*}%
\begin{equation}\fl\qquad
\zeta _{3}(\tau )=\zeta _{3b}+v_{3}\,\left( \tau -\tau _{b}\right)
+\sum\limits_{k=3}^{\infty }\alpha _{3}^{(k)}\,\left( \tau -\tau _{b}\right)
^{k/2},  \label{Locc}
\end{equation}%
with%
\begin{equation}\fl\qquad
\alpha ^{2}=g_{3},~~~v_{3}=-\frac{g_{1}+g_{2}}{\zeta _{b}-\zeta _{3b}}%
,~~~v_{s}=\frac{g_{s}+5\ g_{s+1}}{6\ \left( \zeta _{b}-\zeta _{3b}\right) }
,\quad s=1,2\,\,\mbox{mod}(2),  \label{Locd}
\end{equation}%
and the constants $\alpha _{n}^{(k)}$ determinable (in principle)
recursively (for $k=3,4,...$) by inserting this \textit{ansatz }in (\ref{1}%
), so that, to begin with%
\begin{eqnarray}\fl\qquad\qquad
\alpha _{3}^{(3)} &=&\frac{2\ \alpha \ \left( g_{2}-g_{1}\right) }{3\ \left(
\zeta _{b}-\zeta _{3}\right) ^{2}},  \label{Locf} \\
\fl\qquad\qquad \alpha _{s}^{(3)} &=&-\left( -1\right) ^{s}\frac{\alpha \ }{36\ \left( \zeta
_{b}-\zeta _{3}\right) ^{2}}\left[ 3\ \left( g_{s}-7\ g_{s+1}\right) +\frac{%
\left( g_{1}-g_{2}\right) ^{2}}{g_{3}}\right] ,  \nonumber \\
\fl\qquad\qquad s &=&1,2~~\mbox{mod}(2),  \label{Locg}
\end{eqnarray}%
and so on. The diligent reader will verify the consistency of this
procedure, for any assignment of the \textit{three} constants $\tau
_{b},~\zeta _{b},~$and $\zeta _{3b},$ which remain undetermined\textit{\ }%
except for the obvious restrictions $\tau _{b}\neq 0,$ $\zeta _{b}\neq 0,$ $%
\zeta _{3b}\neq \zeta _{b}.$ The fact that (53) contains \textit{%
three\ }arbitrary (\textit{complex}) constants -- the maximal number of
integration constants compatible with the system of \textit{three\ }%
first-order ODEs (\ref{1}) -- shows that this \textit{ansatz }is indeed
adequate to represent locally, in the neighborhood of its singularities
occurring at $\tau =\tau _{b}$, the \textit{general} solution of (\ref{1}).


An analogous analysis of the behavior of the solutions of the system (\ref{1}%
) near the values of the independent variable $\tau $ where a \textit{triple}
coincidence of all \textit{three} functions $\zeta _{n}$ occurs
(corresponding to the excluded assignment $\zeta _{3b}=\zeta _{b}$ in the
above \textit{ansatz }(53)), indicates, somewhat surprisingly, that
such a \textit{triple} coincidence,
\endnumparts
\begin{equation}
\zeta _{1}(\tau _{b})=\zeta _{2}(\tau _{b})=\zeta _{3}(\tau _{b})=Z
\label{triple}
\end{equation}%
might also occur for the \textit{general }solution of the
system (\ref{1}). This conclusion is reached via a local analysis analogous
to that performed above, and is then confirmed (for the semisymmetrical
case, see (\ref{Symm})) by the \textit{exact }treatment of Section \ref{Sec:General}. Indeed
the natural extension of the above \textit{ansatz} (53)
characterizing the behavior of the solutions of (\ref{1}) in the
neighborhood of such singularities, corresponding to a \textit{triple}
coincidence, see (\ref{triple}), of the \textit{three} functions $\zeta
_{n}(\tau ),$ reads as follows:
\numparts
\begin{equation}\fl\qquad
\zeta _{n}(\tau )=Z+\eta _{n}\,\left( \tau -\tau _{b}\right) ^{\,\left(
1-\gamma \right) /2}+\alpha _{n}\,\left( \tau -\tau _{b}\right)
^{\,1/2}+{\rm o}\left( \left\vert \tau -\tau _{b}\right\vert ^{\,\,1/2}\right) ,
\label{Ansa}
\end{equation}%
provided
\begin{equation}
\mbox{Re}\left( \gamma \right) <0~.  \label{Ansd}
\end{equation}%
Here the \textit{three} constants $\alpha _{n}$ are determined, as can be
easily verified, just by the \textit{three} nonlinear algebraic equations (%
\ref{Ansbb}) that were found in the preceding section while investigating
the equilibrium configurations of our \textit{physical} system (\ref{EqMot}%
), while the \textit{three} constants $\eta _{n},$ as well as the exponent $%
\gamma ,$ are required to satisfy the algebraic equations
\endnumparts
\begin{equation}
\frac{\left( \gamma -1\right) \!\eta _{n}}{2}=\frac{g_{n+1}\!\left( \eta
_{n}-\eta _{n+2}\right) }{\left( \alpha _{n}-\alpha _{n+2}\right) ^{2}}+%
\frac{g_{n+2}\!\left( \eta _{n}-\eta _{n+1}\right) }{\left( \alpha
_{n}-\alpha _{n+1}\right) ^{2}}~.  \label{Ansc}
\end{equation}%
These algebraic equations, (\ref{Ansbb}) and (\ref{Ansc}), can be
conveniently rewritten as follows:
\numparts
\begin{equation}
\alpha _{n}=\beta _{n+1}\,\left( \alpha _{n}-\alpha _{n+2}\right) +\beta
_{n+2}\,\left( \alpha _{n}-\alpha _{n+1}\right) ,  \label{60a}
\end{equation}%
\begin{equation}
\left( \gamma -1\right) \!\eta _{n}=\beta _{n+1}\,\left( \eta _{n}-\eta
_{n+2}\right) +\beta _{n+2}\,\left( \eta _{n}-\eta _{n+1}\right) ,
\label{60c}
\end{equation}%
via the introduction of the quantities $\beta _{n}$, see (\ref{600b}). Note
that in this manner these two sets of equations, (\ref{60a}) and (\ref{60c}%
), have a quite similar look, which should however not mislead the reader to
underestimate their basic difference: the \textit{three} equations (\ref{60a}%
) are merely a convenient way to rewrite, via the definition (\ref{600b}),
the \textit{three} \textit{nonlinear} equations (\ref{Ansbb}), which
determine (albeit not uniquely, see Appendix A) the \textit{three} constants
$\alpha _{n}$; while the equations (\ref{60c}) are \textit{three} \textit{%
linear} equations for the \textit{three} quantities $\eta _{n},$ hence they
can determine these \textit{three} unknowns only up to a common
multiplicative constant (provided they admit a \textit{nontrivial} solution:
see below).

Of course these linear equations (\ref{60c}) admit the \textit{trivial}
solution $\eta _{n}=0,$ and it is easily seen that there indeed is a special
(\textit{exact}) solution of the equations of motion (\ref{1}) having this
property, see (\ref{exact}) with the constants $\alpha _{n}$ determined by (%
\ref{Ansbb}) and computed, for the semisymmetrical model, in Appendix A.
This ``similarity solution'' (\ref{exact})\textit{\ }of the system (\ref{1})
has been discussed in the preceding section; but let us emphasize here that
it only provides a \textit{two}-parameter ($Z$ and $\tau _{b}$) class of
solutions of the equations of motion (\ref{1}), while the \textit{general}
solution of this system of \textit{three} first-order ODEs must of course
feature \textit{three} arbitrary parameters.

A \textit{general }solution of the evolution equations (\ref{1}) corresponds
instead to the \textit{ansatz} (\ref{Ansa}) if the linear equations (\ref{60c}) for the \textit{three} coefficients $\eta _{n}$ admit a \textit{%
nonvanishing} solution, because in such a case, as mentioned above, a common
scaling parameter for these \textit{three} coefficients remains as an
additional (\textit{third}) free parameter (besides $Z$ and $\tau _{b}$).
The condition for this to happen is the vanishing of the determinant of the
coefficients of these \textit{three} linear equations, (\ref{60c}), namely
again validity of the determinantal condition (\ref{Detbeta}), a \textit{%
cubic} equation for the unknown $\gamma ,$ which determines, as discussed in
the preceding section, the \textit{three} values (\ref{3gammas}) of this
quantity. But the first two of these values, $\gamma =\gamma ^{(1)}=1$ and $%
\gamma =\gamma ^{(2)}=2$ (see (\ref{3gammas})), are \textit{not} consistent
with the requirement (\ref{Ansd}). The third solution, $\gamma =\gamma
^{(3)}=2\!\left( \beta _{1}+\beta _{2}+\beta _{3}\right) $ (see (\ref{3gammas})) might instead be consistent with the requirement (\ref{Ansd}),
and whenever this happens the \textit{ansatz }(55) indicates that the
\textit{general }solution of the system of ODEs (\ref{1}) does feature a
``triple coincidence'', see (\ref{triple}), and identifies the character of
the corresponding \textit{branch point}.

In the semisymmetrical case (\ref{Symm}) the equations characterizing the
equilibrium configuration, (\ref{Ansbb}) or equivalently (57), can be
solved (see Appendix A). One finds that there are \textit{two} distinct
solutions of these nonlinear equations (\ref{Ansbb}) (in fact \textit{four},
since each solution has a trivial counterpart obtained by exchanging the
role of the two ``equal'' particles with labels $1$ and $2).$ The \textit{first%
} solution yields for $\gamma =\gamma ^{(3)}$ the value (\ref{Gamma3a}),
which is consistent with the condition (\ref{Ansd}) iff
\endnumparts
\begin{equation}
\mbox{Re}\left( \mu \right) <0,  \label{gamma3ac}
\end{equation}%
and it yields for the branch point exponent, see (\ref{Ansa}), the value
\begin{equation}
\frac{1-\gamma }{2}=\frac{\mu -1}{2\,\mu }=\frac{p-q}{%
2\,p}~;  \label{Expa}
\end{equation}%
while the \textit{second} solution yields for $\gamma ^{(3)}$ the value (\ref{Gamma3b}), which is consistent with the condition (\ref{Ansd}) iff
\begin{equation}
\mbox{Re}\left( \mu \right) >1,  \label{gamma3bc}
\end{equation}%
and it yields for the branch point exponent, see (\ref{Ansa}), the value
\begin{equation}
\frac{1-\gamma }{2}=\frac{\mu +1}{2\,\left( \mu
-1\right) }=\frac{p+q}{2\,(p-q)}~.  \label{Expb}
\end{equation}
The last equality in \eref{Expa} and \eref{Expb} are of course only valid if $\mu$ is rational, $\mu=p/q$.

Note that these findings imply that the branch point associated with ``triple
coincidences'' is not (only) of \textit{square-root} type, being also
characterized, see (\ref{Ansa}), by the exponent $\frac{1-\gamma }{2},$ the
value of which depends on  the value of the parameter $\mu $, see (\ref{Expa}) and (\ref{Expb}); however this kind of branch point is \textit{not} present if
\begin{equation}
0<\mbox{Re}\left( \mu \right) <1,
\label{CondOnlySQBP}
\end{equation}%
since in this case neither (\ref{gamma3ac}) nor (\ref{gamma3bc}) are satisfied.

The results presented in this section are not entirely rigorous, since the
local analysis of the singularities we performed above on the basis of
appropriate \textit{ans\"{a}tze }should be complemented by proofs that the
relevant expansions converge. Moreover these analyses provide information on
the nature of the branch points, but neither on their number nor their
location. But these results are confirmed and complemented below (see
Section \ref{Sec:General}) by the analysis of the \textit{exact} general solution of the
equations of motion (\ref{1}). Our motivation for having nevertheless
presented here a discussion of the character of the singularities of the
solutions of (\ref{1}) via a local analysis \textit{\`{a} la Painlev\'{e} }%
is because an analogous treatment may be applicable to models which are not
as explicitly solvable as that treated in this paper (see for instance \cite
{CS2002} and \cite{CFS2003}).

\bigskip

\section{General solution by quadratures}\label{Sec:General}

In this section we obtain and discuss the \textit{general solutions} of our
models, (\ref{1}) and (\ref{EqMot}). But since the \textit{general solution}
of the \textit{physical} model (\ref{EqMot}) is easily obtained via the
trick (5) from the \textit{general solution} of the \textit{%
auxiliary} problem (\ref{1}), we focus to begin with on this model.

A first constant of the motion is provided by the center-of-mass coordinate
\numparts
\begin{equation}
Z=\frac{\zeta _{1}+\zeta _{2}+\zeta _{3}}{3},  \label{ZZa}
\end{equation}%
since the equations of motion (\ref{1}) clearly entail%
\begin{equation}
Z^{\prime }=0~  \label{ZZb}
\end{equation}%
hence%
\begin{equation}
Z(\tau )=Z(0)~.  \label{ZZc}
\end{equation}%
And clearly the \textit{general solution }of (\ref{1}) reads
\endnumparts
\numparts
\begin{equation}
\zeta _{n}\left( \tau \right) =Z+\check{\zeta}_{n}\left( \tau \right) ,
\end{equation}%
with the set of $3$ functions $\check{\zeta}_{n}\left( \tau \right) $
providing themselves a solution of (\ref{1}), independent of the value of $Z$
and satisfying the (compatible) constraint%
\begin{equation}
\check{\zeta}_{1}\left( \tau \right) +\check{\zeta}_{2}\left( \tau \right) +%
\check{\zeta}_{3}\left( \tau \right) =0~.
\end{equation}%
It is moreover clear that the equations of motion (\ref{1}) entail
\endnumparts
\numparts
\label{Sq}
\begin{equation}
\zeta _{1}^{\prime }\,\zeta _{1}+\zeta _{2}^{\prime }\,\zeta _{2}+\zeta
_{3}^{\prime }\,\zeta _{3}=g_{1}+g_{2}+g_{3},  \label{Sqa}
\end{equation}%
hence there also holds the relation%
\begin{equation}
\zeta _{1}^{2}+\zeta _{2}^{2}+\zeta _{3}^{2}=2\,\left(
g_{1}+g_{2}+g_{3}\right) \,\left( \tau -\tau _{0}\right) ~.  \label{Sqb}
\end{equation}
It is now convenient to set, as in the Appendix B of \cite{CS2002},
\endnumparts
\numparts
\begin{equation}
\zeta _{s}=Z-\left( \frac{2}{3}\right) ^{\,1/2}\,\rho \,\cos \left[ {\theta
-(-1)}^{s}\ \frac{2\ \pi }{3}\right] ,~~~s=1,2,  \label{Sola}
\end{equation}%
\begin{equation}
\zeta _{3}=Z-\left( \frac{2}{3}\right) ^{\,1/2}\,\rho \,\cos \theta ~.
\label{Solc}
\end{equation}%
Then, summing the squares of these three formulas and using the \textit{%
identities}
\endnumparts
\begin{equation}
\cos (\theta )+\cos (\theta +\frac{2\,\pi }{3})+\cos (\theta -\frac{2\,\pi }{%
3})=0,
\end{equation}%
\begin{equation}
\cos ^{2}(\theta )+\cos ^{2}(\theta +\frac{2\,\pi }{3})+\cos ^{2}(\theta -%
\frac{2\,\pi }{3})=\frac{3}{2},
\end{equation}%
one easily gets
\numparts
\begin{equation}
\zeta _{1}^{\,2}+\zeta _{2}^{\,2}+\zeta _{3}^{\,2}=3\ Z^{\,2}+\rho ^{\,2}
\label{Rhoaaa}
\end{equation}%
or equivalently%
\begin{equation}
\rho ^{\,2}=\frac{1}{3}\,\left[ \left( \zeta _{1}-\zeta _{2}\right)
^{\,2}+\left( \zeta _{2}-\zeta _{3}\right) ^{\,2}+\left( \zeta _{3}-\zeta
_{1}\right) ^{\,2}\right] ,  \label{Rhoaab}
\end{equation}%
hence, from (\ref{Sqb}),
\begin{equation}\fl\qquad
\rho ^{2}=2\,\left( g_{1}+g_{2}+g_{3}\right) \,\left( \tau -\tau _{0}\right)
-3\,Z^{2}=2\,\left( g_{1}+g_{2}+g_{3}\right) \,\left( \tau -\tau _{1}\right)
,  \label{Rhoa}
\end{equation}%
\begin{equation}
\tau _{1}=\tau _{0}+\frac{3\,Z^{2}}{2\,\left( g_{1}+g_{2}+g_{3}\right) },
\label{Rhob}
\end{equation}%
which also entails%
\begin{equation}
\rho ^{\prime }\,\rho =g_{1}+g_{2}+g_{3}~.  \label{Rhoc}
\end{equation}%
Here we assume that the sum of the \textit{three} coupling constants $g_{n}$
does not vanish, $g_{1}+g_{2}+g_{3}\neq 0.$ The special case in which this
sum does instead vanish is treated in Appendix C. The expression of the
constant $\tau _{1}$ in terms of the initial data is of course (see (\ref{Rhoa}))
\endnumparts
\numparts
\label{tau1}
\begin{equation}
\tau _{1}=-\frac{\rho ^{2}(0)}{2\,\left( g_{1}+g_{2}+g_{3}\right) },
\label{tau1a}
\end{equation}%
namely (see (\ref{Rhoa}))%
\begin{equation}
\tau _{1}=-\frac{\zeta _{1}^{2}(0)+\zeta _{2}^{2}(0)+\zeta
_{3}^{2}(0)-3\,Z^{2}}{2\,\left( g_{1}+g_{2}+g_{3}\right) },  \label{tau1b}
\end{equation}%
or equivalently (see (\ref{Rhob}))
\begin{equation}
\tau _{1}=-\frac{\left( \zeta _{1}-\zeta _{2}\right) ^{\,2}+\left( \zeta
_{2}-\zeta _{3}\right) ^{\,2}+\left( \zeta _{3}-\zeta _{1}\right) ^{\,2}}{%
6\,\left( g_{1}+g_{2}+g_{3}\right) }~.  \label{tau1c}
\end{equation}
There remains to compute $\theta \left( \tau \right) ,$ or rather
\endnumparts
\begin{equation}
u\left( \tau \right) =\cos \theta \left( \tau \right) ~.  \label{u}
\end{equation}%
Inserting the \textit{ansatz }(66) in the equation of motion (\ref{1}%
) with $n=3,$ one easily gets%
\begin{eqnarray}\fl\qquad
\rho ^{2}\,\left( \cos \theta \right) ^{\prime }\,\left( 4\,\cos
^{2}\theta -1\right) &=&\left( 4\,g_{1}+4\,g_{2}+g_{3}\right) \,\cos \theta -4\,\left(
g_{1}+g_{2}+g_{3}\right) \,\cos ^{3}\theta \nonumber\\
&&\,+\,\sqrt{3}\,\left( g_{1}-g_{2}\right) \,\sin \theta ~.  \label{EqThetaa}
\end{eqnarray}

From now on in this section -- for simplicity, and because it is sufficient
for our purposes -- we restrict attention to the semisymmetrical case (\ref
{Symm}), so that the last equation becomes simply, via (\ref{u}),%
\begin{eqnarray}
\rho ^{2}\,u^{\prime }\,\left( 4\,u^{2}-1\right)  =\left( f+8\,g\right) \,u-4\,\left( f+2\ g\right) \,u^{3}~.
\end{eqnarray}%
The general case \textit{without} the restriction (\ref{Symm}) is treated in
Appendix C.

This ODE can be easily integrated via a quadrature (using (\ref{Rhoa})), and
this leads to the following formula:%
\begin{equation}
\left[ u\left( \tau \right) \right] ^{\!-2\,\mu }\,\left[ u^{2}\left( \tau
\right) -\frac{1}{4\,\mu }\right] ^{\!\mu -1}=K\,\left( \tau -\tau
_{1}\right) ,  \label{xa}
\end{equation}%
where the parameter $\mu $ is defined by (\ref{mu}) and $K$ is an
integration constant. Here we are of course assuming that $f+8\,g\neq 0$
(see (\ref{mu})); the case when this does \textit{not} happen is treated in
Appendix C. (Also recall that, as promised above, we shall treat in Appendix
C the case in which the sum of the three coupling constants $g_{n}$
vanishes, namely when $f+2\,g=0$, which entails $\mu =0,$ see (\ref{mu})).
As for the quantity $K$ in (\ref{xa}), it is an (\textit{a priori arbitrary}%
) integration constant. It is a matter of elementary algebra to express this
constant in terms of the original dependent variables $\zeta _{n}$ (via (\ref
{xa}), (\ref{Rhoa}), (\ref{u}) and (66)), and one thereby obtains the
relation%
\begin{equation}
K=12\,(f+2\,g)\,\tilde{K}  \label{KK}
\end{equation}%
with $\tilde{K}$ defined by (\ref{Ktilde}). This finding justifies the
assertion that $\tilde{K}$ is a constant of motion, see Section 2.3; and of
course it determines the value to be assigned to the constant $K$ in the
context of the \textit{initial-value} problem. Likewise the value to be
assigned, in the context of the \textit{initial-value} problem, to the
constant $\tau _{1}$ appearing in the right-hand side of (\ref{xa}) is given
by the formula%
\begin{equation}
K\,\tau _{1}=-\left[ u\left( 0\right) \right] ^{\!-2\,\mu }\,\left[
u^{2}\left( 0\right) -\frac{1}{4\,\mu }\right] ^{\!\mu -1},  \label{tauone}
\end{equation}%
where (see (\ref{u}) and (\ref{Solc}))
\numparts
\label{u(0)}
\begin{equation}
u(0)=-\left( \frac{3}{2}\right) ^{1/2}\frac{\zeta _{3}(0)-Z}{\rho (0)}
\label{u(0)a}
\end{equation}%
namely
\begin{equation}\fl\qquad
u(0)=-\frac{2\,\zeta _{3}(0)-\zeta _{1}(0)-\zeta _{2}(0)}{\left[ 2\,\left\{ %
\left[ \zeta _{1}(0)-\zeta _{2}(0)\right] ^{\,2}+\left[ \zeta _{2}(0)-\zeta
_{3}(0)\right] ^{\,2}+\left[ \zeta _{3}(0)-\zeta _{1}(0)\right]
^{\,2}\right\} \right] ^{1/2}}~.  \label{u(0)b}
\end{equation}%
Of course in these formulas the initial values $\zeta _{n}(0)$ of the
coordinates $\zeta _{n}(\tau )$ of the \textit{auxiliary} problem (\ref{1})
can be replaced by the initial values $z_{n}(0)$ of the \textit{physical}
problem (\ref{EqMot}), see (\ref{InDat}).

Let us emphasize that we have now reduced, via (66) with (\ref{Rhoa})
and (\ref{u}), the solution of our problem (\ref{1}) to the investigation of
the function $u(\tau )$ of the \textit{complex} variable $\tau ,$ defined
for $\tau \neq 0$ as the solution of the (\textit{nondifferential}) equation
(\ref{xa}) that evolves by continuity from $u(0)$ at $\tau =0$.

To proceed with our analysis an additional change of variables is now
convenient. We introduce the new (\textit{complex}) independent variable $%
\xi $ by setting
\endnumparts
\begin{equation}
\xi =\frac{K\,\left( \tau -\tau _{1}\right) }{4\,\mu },  \label{ksi}
\end{equation}%
and the new (\textit{complex}) dependent variable $w\equiv w\left( \xi
\right) $ by setting
\begin{equation}
w\left( \xi \right) =4\,\mu \,\left[ u\left( \tau \right) \right] ^{\,2}~.
\label{ww}
\end{equation}%
Thereby the expression of the solution (66) of our original problem (%
\ref{1}) reads
\numparts
\begin{eqnarray}\fl\qquad
\zeta _{s}(\tau ) &=&Z-\left( \frac{f+2\,g}{3\,K}\right) ^{\,1/2}\,\xi
^{\,1/2}\,\left\{ -\left[ w\left( \xi \right) \right] ^{\,1/2}+\left(
-\right) ^{s}\,\left[ 12\,\mu -3\,w\left( \xi \right) \right]
^{\,1/2}\right\} ,  \nonumber \\\fl\qquad
~~~s &=&1,2,  \label{SOLUZa}
\end{eqnarray}%
\begin{equation}\fl\qquad
\zeta _{3}\left( \tau \right) =Z-2\,\left( \frac{f+2\,g}{3\,K}\right)
^{\,1/2}\,\left[ \xi \,w\left( \xi \right) \right] ^{\,\,1/2},
\label{SOLUZb}
\end{equation}%
while the (\textit{nondifferential}) equation that determines the dependence
of $w\left( \xi \right) $ on the (\textit{complex}) variable $\xi $ reads
\endnumparts
\begin{equation}
\left[ w\left( \xi \right) -1\right] ^{\,\mu -1}\,\left[ w\left( \xi \right) %
\right] ^{\,-\mu }=\xi ~.  \label{Eqw}
\end{equation}%
Note that this equation is \textit{independent} of the \textit{initial data}%
; it only features the constant $\mu $, which \textit{only} depends on the
coupling constants, see (\ref{mu}).

We conclude that the solution of our \textit{physical} problem (\ref{EqMot})
as the \textit{real} time variable $t$ evolves onwards from $t=0$ is
essentially given, via (80) and (5), by the evolution of
the solution $w(\xi )$ of this (\textit{nondifferential}) equation, (\ref
{Eqw}), as the \textit{complex} variable $\xi $ travels round and round on
the circle $\Xi $ in the \textit{complex }$\xi $-plane defined by the
equation (see (\ref{ksi}) and (\ref{TRICKb}))
\numparts
\begin{equation}
\xi =R\,\exp \left( 2\,i\,\omega \,t\right) +\bar{\xi}=R\,\left[ \exp \left(
2\,i\,\omega \,t\right) +\eta \right] ,  \label{ksiofta}
\end{equation}%
namely on the circle with center $\bar{\xi}$ and radius $\left\vert
R\right\vert .$ The parameters $R$ and $\bar{\xi}$ (or $\eta $) depend on
the initial data according to the formulas (implied by (\ref{ksi}), (5), (\ref{KK}), (\ref{Ktilde}), (\ref{tau1c}))%
\begin{equation}\fl\qquad\quad
R=\frac{3\,\left( f+8\,g\right) }{2\,i\,\omega \,\left[ 2%
\,z_{3}(0)-z_{1}(0)-z_{2}(0)\right] ^{\,2}}\,\left[ 1-\kappa \right] ^{\,\mu
-1},  \label{ksioftd}
\end{equation}%
\begin{equation}\fl\qquad\quad
\bar{\xi}=R\,\eta ,  \label{ksioftg}
\end{equation}%
\begin{equation}\fl\qquad\quad
\eta =\frac{i\,\omega \,\left\{ \left[ \zeta _{1}(0)-\zeta _{2}(0)\right]
^{\,2}+\left[ \zeta _{2}(0)-\zeta _{3}(0)\right] ^{\,2}+\left[ \zeta
_{3}(0)-\zeta _{1}(0)\right] ^{\,2}\right\} }{3\,(f+2\,g)}-1,
\label{ksioftf}
\end{equation}%
\begin{equation}\fl\qquad\quad
\kappa =\frac{2\,\mu \,\left[ 2\,\zeta _{3}(0)-\zeta _{1}(0)-\zeta _{2}(0)%
\right] ^{\,2}}{\left[ \zeta _{1}(0)-\zeta _{2}(0)\right] ^{\,2}+\left[
\zeta _{2}(0)-\zeta _{3}(0)\right] ^{\,2}+\left[ \zeta _{3}(0)-\zeta _{1}(0)%
\right] ^{\,2}}~.  \label{ksiofte}
\end{equation}%
Of course in these formulas the initial values $\zeta _{n}(0)$ of the
coordinates $\zeta _{n}(\tau )$ of the \textit{auxiliary} problem (\ref{1})
can be replaced by the initial values $z_{n}(0)$ of the coordinates $%
z_{n}(t) $ of the \textit{physical} problem (\ref{EqMot}), see (\ref{InDat}).


Let us emphasize that, as the complex variable $\xi $ travels on the circle $%
\Xi $ -- taking the time $T$ to make each round, see (\ref{ksiofta}) and (%
\ref{T}) -- the dependent variable $w\left( \xi \right) $ travels on the
Riemann surface determined by its dependence on the \textit{complex }%
variable $\xi ,$ as entailed by the equation (\ref{Eqw}) that relates $%
w\left( \xi \right) $ to its argument $\xi $ -- starting at $t=0$ from $\xi
=\xi _{0}$,
\endnumparts
\numparts
\label{ksizero}
\begin{equation}\fl\qquad\quad
\xi _{0}=\bar{\xi}+R=\left( \eta +1\right) \,R,  \label{ksizeroa}
\end{equation}%
\begin{equation}\fl\qquad\quad
\xi _{0}=\frac{i\,\omega \,R\,\left\{ \left[ \zeta _{1}(0)-\zeta _{2}(0)%
\right] ^{\,2}+\left[ \zeta _{2}(0)-\zeta _{3}(0)\right] ^{\,2}+\left[ \zeta
_{3}(0)-\zeta _{1}(0)\right] ^{\,2}\right\} }{3\,(f+2\,g)}~  \label{ksizerob}
\end{equation}%
(see (82)) and correspondingly from $w(\xi _{0})=w_{0}$,
\endnumparts
\begin{equation}\fl\qquad\quad
w_{0}=\frac{1}{\kappa }=\frac{\left[ \zeta _{1}(0)-\zeta _{2}(0)\right]
^{\,2}+\left[ \zeta _{2}(0)-\zeta _{3}(0)\right] ^{\,2}+\left[ \zeta
_{3}(0)-\zeta _{1}(0)\right] ^{\,2}}{2\,\mu \,\left[ 2\,\zeta _{3}(0)-\zeta
_{1}(0)-\zeta _{2}(0)\right] ^{\,2}}  \label{ksiwzero}
\end{equation}%
(see (\ref{ksiofte})).

Let us therefore now discuss the structure of this Riemann surface, namely
the analytic properties of the function $w(\xi )$ defined by (\ref{Eqw}).
There are two types of singularities, the ``fixed'' ones occurring at values
of the independent variable $\xi ,$ and correspondingly of the dependent
variable $w,$ that can be read directly from the structure of the equation (%
\ref{Eqw}) under investigation, and the ``movable'' ones (this name being
given to underline their difference from the \textit{fixed} ones) occurring
at values of the independent and dependent variables, $\xi $ and $w,$ that
cannot be directly read from the structure of the equation (\ref{Eqw}) under
investigation (they ``move'' as the initial data are modified).

\subsection{Movable singularities}\label{Sec:movable}

To investigate their nature it is convenient to differentiate (\ref{Eqw}),
obtaining thereby (using again (\ref{Eqw}))
\begin{equation}
\xi \,w^{\prime }=-\frac{w\,\left( w-1\right) }{w-\mu },  \label{EqDiffw}
\end{equation}%
where the prime indicates of course differentiation with respect to $\xi .$
(Note that this ODE is implied by the nondifferential equation (\ref{Eqw}),
while its solution reproduces the nondifferential equation (\ref{Eqw}) up to
multiplication of its right-hand side by an arbitrary constant). The
position of the singularities, $\xi _{b},$ and the corresponding values of
the dependent variable, $w_{b}\equiv w(\xi _{b}),$ are then characterized by
the vanishing of the denominator in the right-hand side of this formula,
yielding the relation%
\begin{equation}
w_{b}=\mu ,  \label{LocSing}
\end{equation}%
which, combined with (\ref{Eqw}) (at $\xi =\xi _{b})$ is easily seen to
yield
\numparts
\begin{equation}
\xi _{b}=\xi _{b}^{(k)}=r\,\exp \left( 2\,\pi \,i\,\mu \,k\right)
,~~~k=1,2,3,...,  \label{Singd}
\end{equation}%
\begin{equation}
\xi _{b}=\xi _{b}^{(k)}=r\,\exp \left[ i\,\frac{2\,\pi \,p\,k}{q}\right]
,~\,~k=1,2,...,q,  \label{Singc}
\end{equation}%
\begin{equation}
r=\left( \mu -1\right) ^{\,-1}\,\left( \frac{\mu -1}{\mu }\right) ^{\,\mu }~.
\label{Singb}
\end{equation}%
In the last, (\ref{Singb}), of these formulas it is understood that the
principal determination is to be taken of the $\mu $-th power appearing in
the right-hand side. The first of these formulas, (\ref{Singd}), shows
clearly that the number of these branch points is \textit{infinite} if the
parameter $\mu $ is \textit{irrational}, and that they then sit densely on
the circle $B$ in the complex $\xi $-plane centered at the origin and having
radius $r$, see (\ref{Singb}). Note that this entails that the \textit{%
generic} point on the circle $B$ is \textit{not} a branch point (just as a
\textit{generic} \textit{real} number is \textit{not rational}); but every
generic point on the circle $B$ has some branch point (in fact, an \textit{%
infinity} of branch points!) \textit{arbitrarily} close to it (just as every
generic \textit{real} number has an \textit{infinity} of \textit{rational}
numbers \textit{arbitrarily} close to it). As for the second of this
formulas, (\ref{Singc}), it is instead appropriate to the case in which the
parameter $\mu $ is \textit{rational}, see (\ref{mupq}), in which case the
branch points sit again on the circle $B$ in the complex $\xi $-plane, but
there are only a \textit{finite} number, $q,$ of them.

These singularities are all \textit{square root }branch points, as implied
by the following standard proof. Set, for $\xi \approx \xi _{b}$,
\endnumparts
\numparts
\label{ansatz}
\begin{equation}
w(\xi )=\mu +a\!\left( \xi -\xi _{b}\right) ^{\beta }+o\left( \left\vert \xi
-\xi _{b}\right\vert ^{\!\mbox{Re}\left( \beta \right) }\right) ,
\label{ansatza}
\end{equation}%
with the assumption (immediately verified, see below) that
\begin{equation}
0<\mbox{Re}\left( \beta \right) <1~.  \label{ansatzb}
\end{equation}%
It is then immediately seen that the insertion of this \textit{ansatz} in (%
\ref{EqDiffw}) (is consistent and) yields
\begin{equation}
\beta =\frac{1}{2},~~~a^{2}=\frac{2\,(1-\mu )}{\xi _{b}}=-2\,\left( \frac{%
\mu }{\mu -1}\right) ^{\,\mu }~.  \label{ansatzc}
\end{equation}%

Note that these results confirm the treatment of Section \ref{Sec:analytic}: the \textit{%
square root} branch points of $w\left( \xi \right) $ identified here, see (%
\ref{LocSing}), are easily seen to correspond, via (80), to the
pair coincidence $\zeta _{1}(\tau _{b})=\zeta _{3}(\tau _{b})$ or $\zeta
_{2}(\tau _{b})=\zeta _{3}(\tau _{b})$; while there is an additional class
of \textit{square-root} branch points which only affect $\zeta _{1}(\tau )$
and $\zeta _{2}(\tau )$, but neither $\zeta _{3}(\tau )$ nor $w\left( \xi
\right) ,$ and occur at
\endnumparts
\numparts
\label{Branch2}
\begin{equation}
w=4\,\mu  \label{Branch2a}
\end{equation}%
due to the vanishing of the second \textit{square-root} term inside the
curly bracket in the right-hand side of (\ref{SOLUZa}), and correspond
therefore to the coincidence $\zeta _{1}(\tau _{b})=\zeta _{2}(\tau _{b})$.
The corresponding values of $\xi $ (as implied by (\ref{Branch2a}) with (\ref
{Eqw})) are%
\begin{equation}
\xi =\frac{\left( 4\,\mu -1\right) ^{\,\mu -1}}{\left( 4\,\mu \right)
^{\,\mu }}=\frac{1}{4\,\mu }\,\left( 1-\frac{1}{4\,\mu }\right) ^{\,\mu -1}~
\label{Branch2b}
\end{equation}%
(we use the plural to refer to these values because of the multivaluedness
of the function in the right-hand side of this formula).

\bigskip

\subsection{Fixed singularities}\label{Sec:fixed}

Next, let us consider the ``fixed'' singularities, which clearly can only
occur at $\xi =\infty $ and at $\xi =0,$ with corresponding values for $w.$

Let us investigate firstly the nature of the singularities at $\xi =\infty .$
Two behaviors of $w(\xi )$ are then possible for $\xi \approx \infty ,$
depending on the value of (the real part of) $\mu $. The first is
characterized by the \textit{ansatz}

\endnumparts
\numparts
\begin{equation}
w(\xi )=a\!\,\xi ^{\!\beta }+o\left( \left\vert \xi \right\vert ^{\!\mbox{Re}%
(\beta )}\right) ,~~~\mbox{Re}(\beta )<0,  \label{Ansatz1a}
\end{equation}%
and its insertion in (\ref{Eqw}) yields%
\begin{equation}
\beta =-\frac{1}{\mu },~~~a^{\mu }=-\exp (i\,\pi \,\mu ),  \label{Ansatz1b}
\end{equation}%
which is consistent with (\ref{Ansatz1a}) iff%
\begin{equation}
\mbox{Re}(\mu )>0~.  \label{Ansatz1c}
\end{equation}%
The second is characterized by the \textit{ansatz}
\endnumparts
\numparts
\label{Ansatz2}
\begin{equation}
w(\xi )=1+a\!\,\xi ^{\!\beta }+o\left( \left\vert \xi \right\vert ^{\!\mbox{Re}\left( \beta \right) }\right) ,~~~\mbox{Re}\left( \beta \right) <0,
\label{Ansatz2a}
\end{equation}%
and its insertion in (\ref{Eqw}) yields%
\begin{equation}
\beta =\frac{1}{\mu -1},~~~a^{\!\mu -1}=1,  \label{Ansatz2b}
\end{equation}%
which is consistent with (\ref{Ansatz2a}) iff%
\begin{equation}
\mbox{Re}(\mu )<1~.  \label{Ansatz2c}
\end{equation}%
We therefore conclude that there are three possibilities: if $\mbox{Re}(\mu
)>1,$ only the first \textit{ansatz}, (90),\textit{\ }is
applicable, and it characterizes the nature of the branch point of $w(\xi )$
at $\xi =\infty ;$ if $\mbox{Re}(\mu )<0,$ only the second \textit{ansatz}, (%
91), is applicable, and it characterizes the nature of the branch
point of $w(\xi )$ at $\xi =\infty ;$ while if $0<\mbox{Re}(\mu )<1,$ both
\textit{ans\"{a}tze}, (90) and (91)\textit{, }are
applicable, so both types of branch points occur at $\xi =\infty .$

Next, let us investigate the nature of the singularity at $\xi =0$. It is
then easily seen, by an analogous treatment, that two behaviors are
possible, as displayed by the following \textit{ans\"{a}tze}: either
\endnumparts
\numparts
\label{Ansatz3}
\begin{equation}
w(\xi )=a\!\,\xi ^{\,\beta }+o\left( \left\vert \xi \right\vert ^{\,\mbox{Re}%
(\beta )}\right) ,~~\ \mbox{Re}(\beta )>0,  \label{Ansatz3a}
\end{equation}%
\begin{equation}
\beta =-\frac{1}{\mu },~~~a^{\,\mu }=-\exp \left( i\,\pi \,\mu \right) ,
\label{Ansatz3b}
\end{equation}%
which is applicable iff%
\begin{equation}
\mbox{Re}(\mu )<0~;  \label{Ansatz3c}
\end{equation}%
or
\endnumparts
\numparts
\label{Ansatz4}
\begin{equation}
w(\xi )=1+a\!\,\xi ^{\,\beta }+o\left( \left\vert \xi \right\vert ^{\,\mbox{%
Re}(\beta )}\right) ,~~\ \mbox{Re}(\beta )>0,  \label{Ansatz4a}
\end{equation}%
\begin{equation}
\beta =\frac{1}{\mu -1},~~~a^{\,\mu -1}=1,  \label{Ansatz4b}
\end{equation}%
which is applicable iff%
\begin{equation}
\mbox{Re}(\mu )>0~.  \label{Ansatz4c}
\end{equation}%
\endnumparts
This analysis shows that the function $w(\xi )$ features a \textit{branch
point} at $\xi =0$ the nature of which is characterized by the relevant
exponent $\beta $, see (\ref{Ansatz3b}) or (\ref{Ansatz4b}), whichever is
applicable (see (\ref{Ansatz3c}) and (\ref{Ansatz4c})). But let us emphasize
that there is \textit{no} branch point \textit{at all} at $\xi =0$ if
neither one of the two inequalities (\ref{Ansatz3c}) and (\ref{Ansatz4c})
holds, namely if $0<\mbox{Re}(\mu )<1$.

\subsection{Explicitly solvable cases}

Let us end this Section \ref{Sec:General} by noting that the equation \eref{Eqw} for certain rational values of $\mu$  reduces to such a low degree polynomial equation that it can be solved \emph{explicitly}. In particular, the polynomial equation is of second degree if $\mu=-1,1/2$ or $2$; it is of third degree if $\mu=-2,-1/2,1/3,2/3,3/2$, or $3$; while it is of fourth degree if $\mu=-3,-1/3,1/4,1/2,3/4,4/3$ or $4$.

The diligent reader might wish to use the corresponding \emph{explicit} solutions formulas for these cases to verify the validity of the previous discussion.

\section{The physical model}\label{Sec:physical}

The solution (80) can also
be written, via (\ref{KK}), (\ref{Ktilde}) and (5), directly for
the particle coordinates $z_{n}(t),$ to read as follows:
\numparts
\label{ZitaSim}
\begin{eqnarray}\fl\quad
z_{s}(t)&=&Z\,{\rm e}^{i\omega t} -\frac{%
2\,z_{3}(0)-z_{1}(0)-z_{2}(0)}{6\,\sqrt{\mu }}\,\left[ \eta \,\exp \left(
-2\,i\,\omega \,t\right) +1\right] ^{\,1/2}\cdot  \nonumber \\
\fl\qquad\quad
&&\cdot \left( -\left[ \check{w}\left( t\right) \right] ^{\,1/2}+\left(
-\right) ^{s}\,\left[ 12\,\mu -3\,\check{w}\left( t\right) \right]
^{\,1/2}\right) ,\qquad s=1,2,  \label{ZitaSima}
\end{eqnarray}%
\begin{equation}\fl\quad
z_{3}\left( t\right) =Z\,{\rm e}^{i\omega t} -\frac{%
2\,z_{3}(0)-z_{1}(0)-z_{2}(0)}{3\,\sqrt{\mu }}\,\left[ \eta \,\exp \left(
-2\,i\,\omega \,t\right) +1\right] ^{\,1/2}\,\left[ \check{w}\left( t\right) %
\right] ^{\,1/2},  \label{ZitaSimb}
\end{equation}%
where the constant $\eta $ is given in terms of the initial data by (\ref
{ksioftf}) and we set
\endnumparts
\begin{equation}
\check{w}(t)\equiv w\left[ \xi \left( t\right) \right] ,  \label{wtilde}
\end{equation}%
so that this dependent variable is now the solution of the nondifferential
equation (see (\ref{Eqw}))%
\begin{equation}\fl\qquad
\left[ \check{w}\left( t\right) -1\right] ^{\,\mu -1}\,\left[ \check{w}%
\left( t\right) \right] ^{\,-\mu }=R\,\exp \left( 2\,i\,\omega \,t\right) +%
\bar{\xi}=R\,\left[ \exp \left( 2\,i\,\omega \,t\right) +\eta \right] ,
\label{Eqwtilde}
\end{equation}%
where the constants $R,$ $\bar{\xi}$ and $\eta $ are defined in terms of the
initial data, see (82) (and recall that the initial data $\zeta
_{n}(0)$ can be replaced by the initial data $z_{n}(0)$, see (\ref{InDat})).
The dependent variable $\check{w}(t)$ is of course the solution of this
equation, (\ref{Eqwtilde}), identified by continuity, as the time $t$
unfolds from $t=0,$ from the initial datum $\check{w}\left( 0\right) =w_{0}$
assigned at $t=0$, see (\ref{ksiwzero}): this specification is necessary,
since generally the nondifferential equation (\ref{Eqw}) has more than a
single solution, in fact possibly even an infinity of solutions.

A discussion of the behaviour of this solution of the initial-value problem of our model \eref{EqMot} with \eref{Symm} clearly hinges on ascertaining how the solution $\check{w}(t)$ of \eref{Eqwtilde} evolves in time. This equation \eref{Eqwtilde} corresponds of course to the combination of \eref{Eqw} with \eref{ksiofta}. Hence one must firstly elucidate the structure of the Riemann surface defined by the dependence of $w(\xi)$ on the complex variables $\xi$ as determined by the nondifferential equations \eref{Eqw}, and then understand the consequences of a travel on this Riemann surface when the complex variable $\xi$ evolves according to \eref{ksiofta}, namely it travels round and round on the circle of center $\bar \xi$ and radius $|R|$ in the complex $\xi$-plane.

The first task is simple, its foundation being provided by the analysis provided in Sections \ref{Sec:movable} and \ref{Sec:fixed}.

The second task is much more demanding, inasmuch as it hinges on the detailed manner the sheets of the Riemann surface are connected via the cuts associated with the branch points discussed in Sections \ref{Sec:movable} and \ref{Sec:fixed}. The main results of this analysis have already been reported (without proofs) in \cite{CGSS2005}; their derivation requires a sufficiently extended treatment to suggest a separate presentation \cite{CGSS2008}. To avoid unnecessary repetitions, also the detailed analysis of the Riemann surface associated to \eref{Eqw}  is postponed to \cite{CGSS2008}.

\section{Outlook}\label{Sec:outlook}

In this paper we report a deeper analysis of the model introduced in \cite{CGSS2005} explaining many results that were there reported without proof (such as the derivation of the general solution by quadratures) while adding some new material (such as a detailed analysis of the equilibrium configurations, small oscillations and similarity solutions of the model).
The novelty of this approach in accounting for a new phenomenology associated to chaotic motion in dynamical systems lies in the fact that the solution is a multi-valued function of (complex) time, and a detailed analysis of its Riemann surface leads to very specific predictions in the simpler ($\mu$ rational) cases,  while it also unveils a source of irregular behaviour in the more complicated ($\mu$ irrational) cases -- unpredictable inasmuch as the determination of its evolution requires knowledge with arbitrarily large precision of the initial data.
The full analysis of the dynamics of this model -- including the geometry of the associated Riemann surface -- is postponed to a future publication \cite{CGSS2008}.

The purpose of this series of papers together with other related projects \cite{FG2006,GS2006}, is to go beyond the local analysis performed in the literature relating analytic properties of solutions in complex time with dynamical properties of the model (Painlev\'e-Kowalewskaya and its non-meromorphic extensions) and to perform a full description of the \emph{global} properties of the Riemann surface. This full description requires not just finding the type of branch points and their positions, but specifying how the different sheets of the Riemann surface are attached together at those branch points.

 Whenever possible, such an approach provides very detailed information on the dynamics that cannot be obtained by the more classical local analyses.

\ack
We would like to thank the Centro Internacional de Ciencias in Cuernavaca, in particular  Fran\c{c}ois Leyvraz and Thomas Seligman, for their support in organizing the \emph{Scientific Gatherings on Integrable Systems and the Transition to Chaos} which provided several opportunities for us to meet and work together.
It is a pleasure to acknowledge illuminating discussions with Boris
Dubrovin, Yuri Fedorov, Jean-Pierre Fran\c{c}oise, Peter Grinevich, Fran\c{c}%
ois Leyvraz, Alexander Mikhailov, Thomas Seligman and Carles Sim\'o.

The research of DGU is supported in part by the Ram\'on y Cajal program of the
Spanish ministry of Science and Technology and by the DGI under grants MTM2006-00478
and MTM2006-14603.

\bigskip

\section*{Appendix A}

In this appendix we solve, in the semisymmetrical case, see (\ref{Symm}),
the nonlinear algebraic equations (\ref{Ansbb}) that characterize the
equilibrium configurations and we thereby compute the ``eigenvalue'' $\gamma
^{(3)}$, namely we obtain its two expressions (\ref{Gamma3a}) and (\ref
{Gamma3b}).

The equations to be solved read (see (\ref{Ansbb}))
\numparts
\label{eqalph}
\begin{equation}
\alpha _{1}=\frac{2\!g}{\alpha _{1}-\alpha _{3}}+\frac{2\!f}{\alpha
_{1}-\alpha _{2}},  \label{eqalpha}
\end{equation}%
\begin{equation}
\alpha _{2}=\frac{2\!g}{\alpha _{2}-\alpha _{3}}-\frac{2\!f}{\alpha
_{1}-\alpha _{2}},  \label{eqalphb}
\end{equation}%
\begin{equation}
\alpha _{3}=\frac{2\!g}{\alpha _{3}-\alpha _{1}}+\frac{2\!g}{\alpha
_{3}-\alpha _{2}},  \label{eqalphc}
\end{equation}%
and they of course imply the relation
\endnumparts
\begin{equation}
\alpha _{1}+\alpha _{2}+\alpha _{3}=0~.
\end{equation}%
It is now convenient to set
\numparts
\label{SD}
\begin{equation}
S=\alpha _{1}+\alpha _{2},~~\ D=\alpha _{1}-\alpha _{2},  \label{SDa}
\end{equation}%
entailing%
\begin{equation}
\alpha _{1}=\frac{S+D}{2},~~~\alpha _{2}=\frac{S-D}{2},~~~\alpha _{3}=-S~.
\label{SDb}
\end{equation}

From (the sum of) (\ref{eqalpha}) and (\ref{eqalphb}) we easily get
\endnumparts
\begin{equation}
S\ \left( 9\ S^{2}-D^{2}\right) =24\ g\ S,
\end{equation}%
and from this we get two types of solutions. The \textit{first} solution is
characterized by $S=0,$ implying (see (\ref{SDb}) and (\ref{eqalpha}))%
\begin{equation}
\alpha _{3}=0,~~~\alpha _{1}=-\alpha _{2}=\alpha ,~~~\alpha ^{2}=f+2\,\!g,
\label{alpha123}
\end{equation}%
entailing (via (\ref{alpha})) the solution (\ref{FirstEqConf}) for the
equilibrium configuration, as well as (via (\ref{600b}) with (\ref{Symm}))
the expressions
\numparts
\begin{eqnarray}
\beta _{3} &=&\frac{f}{2\,\left( f+2\,g\right) },  \label{beta3} \\
\beta _{1} &=&\beta _{2}=\frac{2\ g}{f+2\,g},
\label{beta1}
\end{eqnarray}%
hence, via (\ref{gamma3}), the first expression, (\ref{Gamma3a}), for $%
\gamma ^{(3)}$.

The \textit{second} solution is characterized by
\endnumparts
\begin{equation}
9\ S^{2}-D^{2}=24\ g~.
\end{equation}%
We now subtract (\ref{eqalphb}) from (\ref{eqalpha}) and we thereby easily
get%
\begin{equation}
D^{2}=\frac{-8\ g\ D^{2}}{9\ S^{2}-D^{2}}+4\ f,
\end{equation}%
hence, via the preceding relation,%
\begin{equation}
D^{2}=3\ f,~\ ~S^{2}=\frac{f+8\ g}{3}~.  \label{DS}
\end{equation}%
And via (\ref{600b}) with (\ref{Symm}) and (\ref{SDb}) this is easily seen
to yield%
\begin{equation}
\beta _{1}+\beta _{2}+\beta _{3}=\frac{f+8\ g}{6\ g},
\end{equation}%
namely, via (\ref{gamma3}), the second expression, (\ref{Gamma3b}), of $%
\gamma ^{(3)}.$
Note moreover that, \textit{in both cases}, one gets the relation%
\begin{equation}
\left( \alpha _{1}-\alpha _{2}\right) ^{\,2}+\left( \alpha _{2}-\alpha
_{3}\right) ^{\,2}+\left( \alpha _{3}-\alpha _{1}\right) ^{\,2}=6\,\left(
f+2\,g\right) ,  \label{SumAlph}
\end{equation}%
as can be easily verified from (\ref{alpha123}) as well as from (\ref{SD})
with (\ref{DS}).

\bigskip

\section*{Appendix B}

In this Appendix we consider certain \textit{nongeneric} (classes of)
solutions of our physical problem (\ref{EqMot}), characterized by special
subclasses of initial data.

If the initial data are such that $\eta ,$ hence as well $\bar{\xi},$
\textit{vanish}, $\eta =\bar{\xi}=0$ -- and this entails that the initial
data satisfy the condition%
\begin{equation}\fl\qquad
\left[ z_{1}(0)-z_{2}(0)\right] ^{\,2}+\left[ z_{2}(0)-z_{3}(0)\right]
^{\,2}+\left[ z_{3}(0)-z_{1}(0)\right] ^{\,2}=\frac{3\,(f+2\,g)}{i\,\omega }%
,  \label{etazero}
\end{equation}%
see (82) and (\ref{InDat}); hence these initial data are \textit{%
not} generic, depending only on $2$ arbitrary \textit{complex} parameters
rather than on $3$ such parameters (or, equivalently, only on $1$ rather
than $2$ such parameters besides the trivial constant $Z$ that only affects
the center-of-mass motion, see (\ref{ZitaSim})) -- then the
time evolution of the solution $z_{n}(t)$ of our physical problem (\ref{EqMot}),
see (\ref{ZitaSim}), is clearly periodic with the period $\frac{T}{q}$
rather than $T$. The consequence of this fact are sufficiently obvious not
to require any additional elaboration. An example of this type is that
characterized by the parameters
\numparts
\begin{equation}
\omega =f=2\,\pi ,\quad g=\pi ,\quad\Rightarrow\quad T=\frac{1}{2},\quad \mu =\frac{%
p}{q}=\frac{2}{5},
\end{equation}%
and the initial data%
\[ \fl\qquad z1(0)=0.2,\qquad z_2(0)=-0.79658+0.71779 {\rm i},\qquad z_3(0)=0.59658-0.71779 {\rm i}\]
that imply that the center of mass is initially at the origin and therefore
stays there for all time, $Z=0.$ These initial data are easily seen to satisfy the
condition (\ref{etazero}).

\section*{Appendix C}

In this Appendix we explain how to integrate the ODE (\ref{EqThetaa}) in the
general case when the three coupling constants $g_{n}$ are \textit{all}
different, namely when the restriction (\ref{Symm}) identifying the
semisymmetrical case does \textit{not} apply, and we also provide the
solution of the ODE (\ref{EqThetaa}) in the two special cases (belonging to
the semisymmetrical class characterized by the restriction (\ref{Symm})) the
treatment of which had been omitted in Section \ref{Sec:General}, and as well in another
special case not belonging to the semisymmetrical class.

\bigskip

\subsection*{Solution of equation (\protect\ref{EqThetaa}) in the general case%
}

In this subsection of Appendix C we indicate how the ODE (\ref{EqThetaa})
can be integrated in the general case when the three coupling constants $%
g_{n}$ are \textit{all} different. It is then convenient to set
\endnumparts
\begin{equation}
V\left( \tau \right) =\tan \left[ \theta \left( \tau \right) \right] ,
\end{equation}%
so that this ODE reads%
\begin{equation}
\frac{V^{\prime }\,V\,\left( V^{2}-3\right) }{\left( V^{2}+1\right) \,\left(
A\,V^{3}+C\,V^{2}+A\,V+C-2\right) }=\frac{1}{\left( \tau -\tau _{1}\right) }
\label{ODE}
\end{equation}%
with%
\begin{equation}
A=\frac{\sqrt{3}\,\left( g_{1}-g_{2}\right) }{2\,\left(
g_{1}+g_{2}+g_{3}\right) },~~~C=\frac{4\,g_{1}+4\,g_{2}+g_{3}}{2\,\left(
g_{1}+g_{2}+g_{3}\right) }~.
\end{equation}
To integrate this ODE we set%
\begin{equation}\fl\qquad
A\,V^{3}+C\,V^{2}+A\,V+C-2=A\,\left( V-V_{1}\right) \,\left( V-V_{2}\right)
\,\left( V-V_{3}\right) ,
\end{equation}%
so that the \textit{three} quantities $V_{n}$ are the \textit{three} roots
of this polynomial of \textit{third} degree in $V.$ We then decompose this
rational function of $V$ in simple fractions,%
\begin{equation}
\frac{V\,\left( V^{2}-3\right) }{\left( V^{2}+1\right) \,\left(
A\,V^{3}+C\,V^{2}+A\,V+C-2\right) }=\sum\limits_{j=1}^{5}\frac{\mu _{j}}{%
V-V_{j}},  \label{VVV}
\end{equation}%
where of course%
\begin{equation}
V_{4}=i,~~~V_{5}=-i,  \label{V45}
\end{equation}%
and the \textit{five} quantities $\mu _{j}$ are easily evaluated in
terms of the $3$ roots $V_{n}$:
\begin{equation}%
\mu_{j}=V_{j}\,(-3+V_{j}^{2})\,\prod_{k=1,k\neq
j}^{5}(V_{j}-V_{k})^{-1} \,\,.
\end{equation}%
The integration of the ODE (\ref{ODE}) is now trivial (using
(\ref{VVV})),
and it yields (using (\ref{V45})) the final formula%
\begin{equation}
\left[ V\left( \tau \right) -i\right] ^{\mu _{4}}\,\left[ V\left(
\tau
\right) +i\right] ^{\mu _{5}}\prod\limits_{n=1}^{3}\left[ V(\tau )-V_{n}%
\right] ^{\mu _{n}}=K\,\left( \tau -\tau _{1}\right) ,
\end{equation}%
where $K$ is the integration constant.

\bigskip

\subsection*{Solution of equation(\protect\ref{EqThetaa}) in two special
subcases of the semisymmetrical case}

In this subsection of Appendix C we provide the solution of the ODE (\ref
{EqThetaa}) in the two special subcases (of the semisymmetrical case) the
treatment of which had been omitted in Section \ref{Sec:General}, and as well in another
special case \textit{not} belonging to the semisymmetrical class.
If
\begin{equation}
g_{1}+g_{2}+g_{3}=0,  \label{SpCas1}
\end{equation}%
$\rho $ is constant (namely $\tau $-independent, $\rho \left( \tau \right)
=\rho \left( 0\right) ,$ see (\ref{Rhoa})). Moreover, via the restriction (%
\ref{Symm}) characterizing the semisymmetrical class, we get (see also (\ref
{mu}))
\numparts
\label{SpCase1}
\begin{equation}
f=-2\,g,~~~\mu =0~.  \label{SpCase1a}
\end{equation}%
Then (\ref{xa}) is replaced by%
\begin{equation}
u\left( \tau \right) \,\exp \left[ -2\,u^{2}\left( \tau \right) \right]
=\exp \left[ \frac{3\,f\,\left( \tau -\tau _{0}\right) }{\rho ^{2}\left(
0\right) }\right] ~.  \label{SpCase1b}
\end{equation}%
\qquad

Let us also note that, if (\ref{Symm}) were replaced by
\endnumparts
\numparts
\label{SpCase11}
\begin{equation}
g_{1}=-g_{2}=g,~~~~g_{3}=0,  \label{SpCas11a}
\end{equation}%
which is also consistent with (\ref{SpCas1}), then (\ref{xa}) with (\ref{u})
would be replaced by
\begin{equation}
\theta \left( \tau \right) +\sin \left[ 2\,\theta \left( \tau \right) \right]
=\frac{2\,\sqrt{3}\,g\,\left( \tau _{0}-\tau \right) }{\rho ^{2}\left(
0\right) }~.  \label{SpCase11b}
\end{equation}
Returning to the semisymmetrical case characterized by validity of the
restriction (\ref{Symm}) we now consider the second case whose treatment had
been omitted in Section \ref{Sec:General}, namely
\endnumparts
\label{SpCase2}
\begin{equation}
f=-8\,g~.  \label{SpCase2a}
\end{equation}%
Note that in this case $\mu $ diverges, see (\ref{mu}). Then (\ref{xa}) is
replaced by%
\begin{equation}
u\left( \tau \right) \,\exp \left[ u^{2}\left( \tau \right) \right] =\left[
K\,\left( \tau -\tau _{1}\right) \right] ^{\!-1/2}~.  \label{SpCase2b}
\end{equation}

\section*{Appendix D: relation with more standard (Newtonian) three-body
problems}

In this Appendix we indicate the relation among the three-body problems
treated in this paper, characterized by equations of motion of \textit{%
Aristotelian }type (``the particle \textit{velocities} are proportional to
assigned external and interparticle forces''), with analogous many-body
problems characterized by equations of motion of \textit{Newtonian} type
(``the particle \textit{accelerations} are proportional to assigned external
and interparticle forces''). The results reviewed in this section are of
interest inasmuch as they relate the model treated in this paper to other,
somewhat more physical and certainly more classical, many-body problems,
including a prototypical three-body model introduced, and shown to be
solvable by quadratures, by Carl Jacobi one and a half centuries ago \cite%
{Jac}, and the one-dimensional Newtonian many-body problem with two-body
forces proportional to the inverse cube of the interparticle distance
introduced and solved over four decades ago (firstly in the quantal context
\cite{C1969,C1971} and then in the classical context \cite{Mar,Mos}), which contributed to the bloom in the investigation of integrable
dynamical systems of the last few decades (see for instance \cite{perelomov,C2001}).

By differentiating the equations of motion (\ref{1}) and using them again to
eliminate the first derivatives in the right-hand sides one gets the
following \textit{second-order} equations of motion of Newtonian type:
\begin{eqnarray}\fl\qquad
\zeta _{n}^{\prime \prime } &=&-\frac{2\,g_{n+1}^{2}}{\left( \zeta
_{n}-\zeta _{n+2}\right) ^{3}}-\frac{2\,g_{n+2}^{2}}{\left( \zeta _{n}-\zeta
_{n+1}\right) ^{3}}  \nonumber \\
\fl\qquad
&&+\frac{g_{n+1}\,\left( g_{n}-g_{n+2}\right) }{\left( \zeta _{n}-\zeta
_{n+2}\right) ^{2}\,\left( \zeta _{n+2}-\zeta _{n+1}\right) }+\frac{%
g_{n+2}\,\left( g_{n}-g_{n+1}\right) }{\left( \zeta _{n}-\zeta _{n+1}\right)
^{2}\,\left( \zeta _{n+1}-\zeta _{n+2}\right) }~.  \label{53}
\end{eqnarray}%
Likewise from the equations of motion (\ref{EqMot}) one gets%
\begin{eqnarray}\fl\qquad
\ddot{z}_{n}+\omega ^{2}\,z_{n} &=&-\frac{2\,g_{n+1}^{2}}{\left(
z_{n}-z_{n+2}\right) ^{3}}-\frac{2\,g_{n+2}^{2}}{\left( z_{n}-z_{n+1}\right)
^{3}}  \nonumber \\
\fl\qquad
&&+\frac{g_{n+1}\,\left( g_{n}-g_{n+2}\right) }{\left( z_{n}-z_{n+2}\right)
^{2}\,\left( z_{n+2}-z_{n+1}\right) }+\frac{g_{n+2}\,\left(
g_{n}-g_{n+1}\right) }{\left( z_{n}-z_{n+1}\right) ^{2}\,\left(
z_{n+1}-z_{n+2}\right) }~.  \nonumber \\
&&  \label{54}
\end{eqnarray}%
Of course the solutions of the \textit{first-order} equations of motion, (%
\ref{1}) respectively (\ref{EqMot}), satisfy as well the corresponding
\textit{second-order} equations of motion, (\ref{53}) respectively (\ref{54}%
), but they provide only a subset of the solutions of the latter. On the
other hand it is again true that the solutions of the second-order equations
of motion (\ref{53}) and (\ref{54}) are related via the trick.

In the integrable ``equal-particle'' case, see (\ref{Integr}), these equations
of motion simplify and correspond respectively to the Newtonian equations of
motion yielded by the two standard $N$-body Hamiltonians%
\begin{equation}
H\left( \underline{\zeta },\,\underline{\pi }\right) =\sum\limits_{n=1}^{N}%
\frac{\pi _{n}^{2}}{2}-\sum\limits_{m,n=1;\,m\neq n}^{N}\frac{g^{2}}{%
2\,\left( \zeta _{n}-\zeta _{m}\right) ^{\,2}},
\end{equation}%
respectively%
\begin{equation}
H\left( \underline{z},\,\underline{p}\right) =\sum\limits_{n=1}^{N}\frac{%
p_{n}^{2}+\omega ^{2}\,z_{n}^{2}}{2}-\sum\limits_{m,n=1;\,m\neq n}^{N}\frac{%
g^{2}}{2\,\left( z_{n}-z_{m}\right) ^{\,2}},
\end{equation}%
with $N=3,$ the complete integrability of which is by now a classical result
(even in the $N$-body case with $N>3$: see for instance \cite{C2001}).

In fact the more general \textit{three-body }Hamiltonian models%
\begin{equation}
H\left( \underline{\zeta },\,\underline{\pi }\right) =\sum\limits_{n=1}^{3}%
\left[ \frac{\pi _{n}^{2}}{2}-\frac{g_{n}^{2}}{\left( \zeta _{n+1}-\zeta
_{n+2}\right) ^{2}}\right] ,
\end{equation}%
respectively%
\begin{equation}
H\left( \underline{z},\,\underline{p}\right) =\sum\limits_{n=1}^{3}\left[
\frac{p_{n}^{2}+\omega ^{2}\,z_{n}^{2}}{2}-\frac{g_{n}^{2}}{\left(
z_{n+1}-z_{n+2}\right) ^{2}}\right] ,
\end{equation}%
featuring three \textit{different} coupling constants $g_{n},$ that yield
the equations of motion%
\begin{equation}
\zeta _{n}^{\prime \prime }=-\frac{2\,g_{n+1}^{2}}{\left( \zeta _{n}-\zeta
_{n+2}\right) ^{3}}-\frac{2\,g_{n+2}^{2}}{\left( \zeta _{n}-\zeta
_{n+1}\right) ^{3}},  \label{51}
\end{equation}%
respectively%
\begin{equation}
\ddot{z}_{n}+\omega ^{2}\,z_{n}=-\frac{2\,g_{n+1}^{2}}{\left(
z_{n}-z_{n+2}\right) ^{3}}-\frac{2\,g_{n+2}^{2}}{\left( z_{n}-z_{n+1}\right)
^{3}},  \label{52}
\end{equation}%
are also solvable by quadratures. For the equations of motion (\ref{51})
this discovery is due to Carl Jacobi \cite{Jac}; while the solutions of the
equations of motion (\ref{52}) can be easily obtained from those of the
equations of motion (\ref{51}) via the trick (5). For a detailed
discussion of these solutions, and additional indications on key
contributions to the study of this problem, the interested reader is
referred to \cite{CS2002} and \cite{C2001}. But we will perhaps also revisit this
problem, because we believe that additional study of these models, (\ref{51}%
) and (\ref{52}), shall shed additional light on the mechanism responsible
for the \textit{onset} of a certain kind of \textit{deterministic chaos}, as
discussed above.

Finally let us recall that our Aristotelian model \eref{EqMot}, as well as the Newtonian models described in this Appendix, describing ``particles'' moving in the \emph{complex} $z$-plane, can be easily reformulated as models describing particles moving in the \emph{real} plane, with \emph{rotation-invariant} (or at least \emph{covariant}) \emph{real} two-vector equations of motion (see for instance Chapter 4 of \cite{C2001}).
%

\end{document}